\documentclass[aps, prl, reprint, superscriptaddress]{revtex4-2}
\usepackage{bbold}
\usepackage{comment}
\usepackage[psdextra, colorlinks, citecolor=red]{hyperref}
\usepackage{hyperref}
\usepackage[utf8]{inputenc}
\usepackage[T1]{fontenc}   
\usepackage[english]{babel}
\usepackage[pdftex]{graphicx}
\usepackage{amsmath, amssymb, amsthm, bbm, mathtools} 
\usepackage{mathrsfs}
\usepackage{xcolor}
\usepackage[normalem]{ulem}
\usepackage{multirow}
\usepackage{braket}
\usepackage{upgreek}
\usepackage{threeparttable}
\usepackage{multirow}
\usepackage{booktabs}

\newcommand{\abs}[1]{\left\vert#1\right\vert}

\begin{document}

\title{Strain engineering of Andreev spin qubits in Germanium}
\author{Vittorio Coppini}
\email{vittorio.coppini@uniroma1.it}
\affiliation{Dipartimento di Fisica, Sapienza Università di Roma, Piazzale Aldo Moro 2, 00185 Rome, Italy}
\author{Patrick Del Vecchio}
\email{p.delvecchio@tudelft.nl}
\affiliation{QuTech and Kavli Institute of Nanoscience, Delft University of Technology, Delft, Netherlands}
\author{Antonio L. R. Manesco}
\affiliation{Kavli Institute of Nanoscience, Delft University of Technology, Delft, Netherlands}
\affiliation{Center for Quantum Devices, Niels Bohr Institute, University of Copenhagen, DK-2100 Copenhagen, Denmark}
\author{Anton Akhmerov}
\affiliation{Kavli Institute of Nanoscience, Delft University of Technology, Delft, Netherlands}
\author{Valla Fatemi}
\affiliation{School of Applied and Engineering Physics, Cornell University, Ithaca, NY, 14853, USA
}
\author{Bernard van Heck}
\affiliation{Dipartimento di Fisica, Sapienza Università di Roma, Piazzale Aldo Moro 2, 00185 Rome, Italy}
\author{Stefano Bosco}
\affiliation{QuTech and Kavli Institute of Nanoscience, Delft University of Technology, Delft, Netherlands}

\begin{abstract}
Planar germanium heterostructures are promising hosts for hybrid quantum devices due to their compatibility with superconductors, low material disorder, and relaxed fabrication constraints.
Also, the potentially low density of nuclear spins and strong spin–orbit interaction make germanium attractive for coherent spin physics. 
However, recent microwave spectroscopy experiments were unable to resolve a spin-splitting of bound states in germanium Josephson junctions, the prerequisite for defining and controlling Andreev spin qubits. 
Here, we argue that compressive strain is the key mechanism  suppressing spin splitting in current devices. 
Furthermore, we propose unstrained and tensile-strained heterostructures, fully compatible with state-of-the-art growth technology, that significantly enhance the relevant spin–orbit effect. 
By numerically simulating ballistic Josephson junctions, we predict spin splittings comfortably in the GHz range, more than 2 orders of magnitude larger than compressively strained cases, and all-electric quantum gates in a hundred nanoseconds.
Our results establish strain engineering as a key design principle for realizing Andreev spin qubits in germanium-based devices.
\end{abstract}

\maketitle

\footnotetext[1]{Supplemental Material containing details on the $k\cdot p$ effective Hamiltonian, on the material parametrization, and on the model of the ASQ.}

\paragraph{Introduction. --}
\begin{figure}
    \centering
    \includegraphics[width=\linewidth]{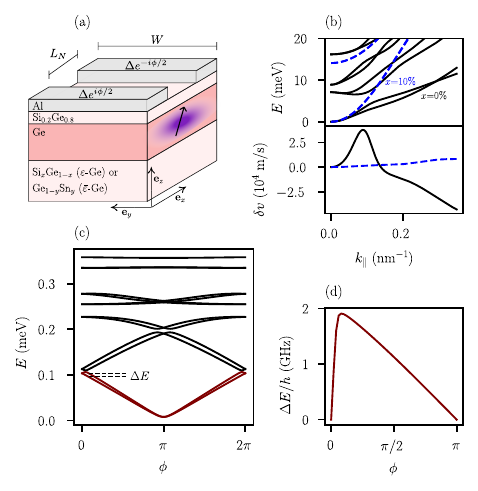}
    \caption{\emph{ASQs in Ge.} (a) Ge Josephson junction of length $L_N$. Two aluminum leads with gap $\Delta$, phase-difference $\phi$, and width $W$, are coupled to a Ge heterostructure comprising a $10$~nm Si$_{0.2}$Ge$_{0.8}$ barrier, a $15$~nm compressive-strained $\varepsilon$-Ge (tensile-strained $\bar\varepsilon$-Ge) channel, and a bottom relaxed barrier of Si$_x$Ge$_{1-x}$ (Ge$_{1-y}$Sn$_y$). At $x=y=0$, the Ge channel is unstrained and accumulated by the electric field $E_z=1$~mV/nm. {(b)} Band dispersion (upper panel) of $\varepsilon$-Ge at $x=2\%$ (black) and $x=10\%$ (blue). As $x$ (i.e., compressive strain) increases, the SOI-induced difference of Fermi velocities $\delta v$ of the two spins substantially decreases (lower panel). {(c)} $\phi$-dependent energy of  Andreev levels in unstrained Ge junctions ($x=y=0$). We highlight in red the lowest spin doublet, whose energy difference $\Delta E$ is shown in {(d).}    }
    \label{fig:fig1}
\end{figure}

Andreev spin qubits (ASQs) are encoded in the spin of a quasiparticle trapped in an Andreev bound state of a superconducting-semiconducting Josephson junction ~\cite{chtchelkatchev2003,padurariu2010,park2017}.
ASQs hold promise to combine the compactness and high-coherence operation of semiconductor spin qubits~\cite{burkard2023} with the long-range coupling and fast microwave readout of superconducting circuits~\cite{blais2021}.
A key requirement for ASQs is a sufficiently strong spin-orbit interaction (SOI) in the semiconductor, which lifts the spin degeneracy of Andreev levels and couples the spin to the supercurrent, enabling qubit definition and control~\cite{chtchelkatchev2003,padurariu2010,park2017,beri2006,dellanna2008,michelsen2008,reynoso2012,tosi2019,hoffman2025}.

So far, proof-of-principle experiments~\cite{tosi2019,hays2020,hays2021,pita2023,bargerbos2023,pita2024} have mainly focused on indium-arsenide (InAs) nanowires, which present strong SOI but are challenging to scale and suffer from intrinsically limited spin coherence due to hyperfine noise caused by nuclear spins~\cite{stano2022}.
In contrast, quantum devices based on planar germanium (Ge) heterostructures offer scalable fabrication, compatible with semiconductor industry. Moreover, Ge can be isotopically purified~\cite{https://doi.org/10.1002/adma.202305703} reducing hyperfine noise~\cite{PhysRevB.78.155329,PhysRevB.101.115302,PhysRevLett.127.190501}, and has emerged as a leading platform for quantum computing with quantum dot spin qubits~\cite{scappucci2021,Borsoi2024,John2025,Hendrickx2021,dijkema2026simultaneous,Zhang2025}.
Progress in achieving high-quality superconducting contacts and proximity-induced gaps further highlights planar Ge as a promising scalable alternative for hybrid quantum devices~\cite{hoffman2025ge,Hendrickx2018,Sagi2024,kiyooka2026andreev,Kiyooka2025,Leblanc2025,fabris2026granular,Lakic2025,Tosato2023,PhysRevResearch.3.L022005,hinderling2024,tenkate2025,Steele2025}.
However, recent experiments on hybrid devices defined in compressively-strained Ge channels ($\varepsilon$-Ge) confined by relaxed silicon-germanium (SiGe) barriers have not observed sizeable spin splitting of Andreev levels, hindering the realization of ASQs~\cite{hinderling2024,tenkate2025}.

In this work, we show that compressive strain suppresses the spin splitting, consistent with current experiments.
By numerically simulating the spin–orbit-induced splitting of Andreev levels in ballistic Josephson junctions with the geometry of Fig.~\ref{fig:fig1}(a), we identify alternative Ge heterostructures fully-compatible with state-of-the-art material growth techniques.
In particular, we find that recently realized unstrained Ge channels sandwiched between strained SiGe barriers~\cite{costa2025} not only provide a lattice-matched, defect-free host material, but they also enable spin splittings in the GHz range, two orders of magnitude larger than in compressively strained Ge.
Moreover, we estimate that the spin splitting can be further enhanced in tensile-strained Ge channels ($\bar\varepsilon$-Ge) confined by relaxed germanium-tin (GeSn) barriers~\cite{Kaul2025}, reaching values of several GHz.
Our results identify strain engineering as a key ingredient for designing ASQ devices in germanium and provide a pathway toward integrating ASQs into the germanium quantum information route.

\paragraph{Effective model of Ge. --} 
Holes in Ge heterostructures are accurately described by the Hamiltonian
\begin{equation}\label{eq.HGe}
H_\text{Ge}=H_{\text{LK}}+H_{\text{BP}}-e E_z z,
\end{equation}
\noindent where $H_{\text{LK}}$ and $H_\text{BP}$ are the 6-band isotropic Luttinger–Kohn~\cite{Eissfeller2011,Winkler2003} and Bir–Pikus Hamiltonian~\cite{Bir1974}, which describe the mixing of heavy holes (HH), light holes (LH), and split-off holes at the $\Gamma$ point induced by the crystal momentum $\textbf{k}$ and by the strain tensor $\varepsilon_{ij}$, respectively.
See the Supplemental Material (SM)~\cite{Note1} for more details on $H_\mathrm{Ge}$~\cite{Winkler2003,VandeWalle1986,VandeWalle1989,Reeber1996,Polak2017,Lawaetz1971,Madelung1991,LuLow2012} and  the dependence of strain $\varepsilon_\parallel=\varepsilon_{xx}=\varepsilon_{yy}$ on the Si ($x$) and Sn ($y$) concentrations.
We include an electric field $E_z = 1\,\mathrm{mV}/\mathrm{nm}$ perpendicular to the substrate,  required to accumulate holes, tune the chemical potential $\mu$ of the channel, and generate Rashba SOI by breaking inversion symmetry.
We restrict ourselves to the analysis of homogeneous strain arising from the mismatch of lattice constants of channel and barriers.
This contribution energetically favors  LHs in the channel in tensile-strained $\bar\varepsilon$-Ge~\cite{PhysRevB.110.045409,del2025fully,PhysRevB.107.L161406} and HHs in compressively-strained $\varepsilon$-Ge~\cite{terrazos2021,Winkler2003}.
HHs are also favored by confinement~\cite{PhysRevB.104.115425,Mauro2025} and remain the ground state in relaxed unstrained Ge heterojunctions, where  the channel is defined by the electric field $E_z$~\cite{costa2025}.
The lack of strain reduces the HH-LH energy gap, strongly enhancing SOI~\cite{PhysRevB.104.115425,Mauro2025}.

Following Refs.~\cite{Winkler2003,PhysRevB.110.045409}, we first find the sub-band states confined in the $z$-direction, perpendicular to the substrate, at $\mathbf{k}_\parallel=(k_x,k_y)=\mathbf{0}$ by diagonalizing Eq.~\eqref{eq.HGe} for different heterostructures.
Using these states, we construct an effective theory for the long-wavelength in-plane dynamics by a Schrieffer-Wolff transformation~\cite{Bravyi2011,Winkler2003}. 

Ge/SiGe heterostructures across the full range of compressive strain, including unstrained Ge channels, are accurately modeled by a low-energy $14\times 14$ effective Hamiltonian ${H}_\mathrm{eff}(\mathbf{k}_\parallel)$
\begin{equation}\label{Heff}
    {H}_\mathrm{eff} = \mathcal{E}_0+\frac{\hbar^2}{2m_0}\left[\mathcal{M}_\gamma k_\parallel^2 + \left(i\mathcal{M}_1k_- + \mathcal{M}_2 k_-^2 + \text{h.c.}\right)\right],
\end{equation}
quadratic in $\mathbf{k}_\parallel$ and describing the planar motion of the first $5$ HH spin doublets and the first $2$ LH composite spin doublets.
Here, $k_\pm = k_x \pm ik_y$, $k_\parallel=|\textbf{k}_\parallel|$, $\mathcal{E}_0$ is the $14\times 14$ diagonal matrix containing the orbital sub-band energies at $\mathbf{k}_\parallel = 0$, and the $14\times 14$ $\mathcal{M}$-matrices describe spin-dependent effective mass and SOI~\cite{Note1}.

\begin{figure*}[t!]
    \includegraphics[width=\textwidth]{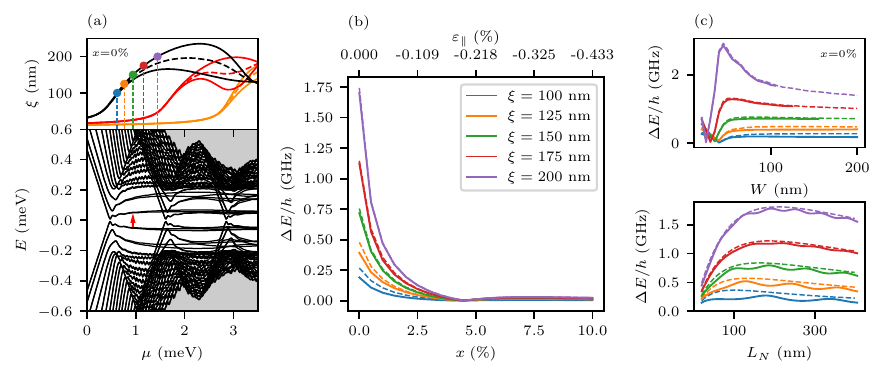}
    \caption{\emph{Andreev spin splitting in $\varepsilon$-Ge.} (a) Coherence lengths $\xi$ (upper panel) and energy levels (lower panel) against $\mu$ in an unstrained Ge junction. We show $\xi$ of different spins (solid lines), and their average (dashed lines) for different sub-bands. The dots mark the values of the $\xi$ used in {(b)} and {(c)}; the arrow marks the value of $\mu$ in Figs.~\ref{fig:fig1}(c)-(d). {(b)} The spin-splitting $\Delta E$ rapidly decreases from the GHz range for unstrained Ge as the Si concentration $x$ in the barrier, which determines the compressive strain of Ge, increases. Solid and dashed lines correspond to numerical simulations and  Eq.~\eqref{eq:transcendent_ABS_equation}, respectively. In panels (a)-(b) we use $L_N=300$ nm, $W=100$ nm, $\phi=\pi/2$. {(c)} Dependence of $\Delta E$ at $\phi=\pi/2$ in unstrained Ge ($x=0$) on the junction width $W$ at fixed length $L_N=300$~nm (upper panel) and on $L_N$ at fixed $W=100$~nm (lower panel). For a wide range of junction geometry and $\xi$s, $\Delta E$ remains comfortably in the GHz range.}
    \label{fig:spin_splitting}
\end{figure*}

Fig.~\ref{fig:fig1}(b) shows the energy dispersion of the low-energy bands in $\varepsilon$-Ge for different strain values, determined by the Si content $x$ in the barrier.
As the strain decreases (lower $x$), the SOI increases, leading to a larger momentum-dependent energy spitting between spin up and down.
By plotting the Fermi velocity difference $\delta v$ between different spin orientations at the same $k_\parallel$, we also note that the SOI exhibits a non-monotonic dependence on momentum, reaching a maximum when the lowest HH band anti-crosses with the first LH band.
A large $\delta v$ is a key requirement for generating sizeable spin splitting of ASQs at zero Zeeman fields.
For Rashba nanowires, this arises due to the interplay of confinement and SOI for 1D sub-bands~\cite{park2017,tosi2019}; here, we see that the SOI of unstrained Ge provides this effect already at the level of the 2-dimensional hole band structure. 

\paragraph{ Ge Josephson Junction. --} We assume that $s$-wave superconductivity is induced in the Ge channel from the superconductor deposited above the barrier.
Starting from the effective Hamiltonian of the Ge channel in Eq.~\eqref{Heff}, the Josephson junction of Fig.~\ref{fig:fig1}(a) is modeled by the  Bogoliubov-de Gennes Hamiltonian
\begin{equation} \label{eq:H-bogo}
\mathcal{H}_\textrm{BdG} =
\begin{bmatrix}
H_\mathrm{eff}-\mu & H_\Delta \\ -H_\Delta^* & -H^*_\mathrm{eff}+\mu
\end{bmatrix}
\end{equation}
where $\mu$ is the chemical potential and $H_\Delta$ describes the induced $s$-wave pairing, characterized by an induced gap parameter $\Delta$.
We choose $\Delta=0.2$~meV with aluminum in mind as the parent superconductor~\cite{Hendrickx2018,Sagi2024,kiyooka2026andreev,Kiyooka2025,Leblanc2025}, while the phase of the induced order parameter is assigned as shown in Fig.~\ref{fig:fig1} to give a phase difference $\phi$ across the junction.
We consider the superconductor equally coupled to the sub-bands of $H_\text{eff}$ and neglect differences in couplings to HHs and LHs~\cite{mmsd-wfnf,k4jh-pnxy,PhysRevB.89.184507,dqgc-8crs,PhysRevB.108.155433}.
Within this minimal model, alternative superconducting material choices such as
niobium~\cite{PhysRevResearch.3.L022005}, granular aluminum~\cite{fabris2026granular}, or platinum-germanium-silicides~\cite{Lakic2025,Tosato2023} only affect the value of the induced gap $\Delta$.
We discretize $\mathcal{H}_\mathrm{BdG}$ on a finite system using Kwant~\cite{kwant} and compute the energy spectrum of the junction numerically. In the simulations, the junction width $W$ is imposed via hard-wall boundary conditions, and the junction length $L_N$ by setting $H_\Delta=0$ in the junction region~\cite{Note1}.

The typical low-energy spectrum of the junction against the phase difference $\phi$ is shown in Fig.~\ref{fig:fig1}(c). We focus on the lowest two Andreev levels, which form a Kramers doublet at $\phi=0$ and $\phi=\pi$. At other values of $\phi,$ their energy difference $\Delta E$, shown in Fig.~\ref{fig:fig1}(d), is directly related to the SOI-induced Fermi-velocity difference $\delta v$~\cite{Note1}. We now investigate the magnitude of $\Delta E$ and its dependence on strain and junction design.

\paragraph{Andreev spin splitting in $\varepsilon$-Ge. --} We start by investigating the spin splitting in Ge/Si$_x$Ge$_{1-x}$ heterostructures.  Figure~\ref{fig:spin_splitting}(a) shows the dependence of the Andreev levels of an unstrained channel ($x=0$) at $\phi=\pi/2$ against the chemical potential $\mu$.
We also show the $\mu$-dependence of the coherence length $\xi$ of evanescent modes in the superconducting leads.

Increasing $\mu$, multiple sub-bands become occupied in succession, as revealed by near-closings of the gap in the energy spectrum and by the onset of extra modes with a large coherence length in the leads.
We note that $\xi$  inherits the spin-dependence of the Fermi velocity of the corresponding sub-bands.
We focus on the range of $\mu$ where only a single band is occupied, such that the low-energy spectrum consists of a well-isolated Andreev doublet encoding the qubit.

As the Si concentration $x$ increases, Ge is compressively strained and the spin splitting rapidly decreases [Fig.~\ref{fig:spin_splitting}(b)].
To compare different $\varepsilon$-Ge channels and junction designs,  we determine five values of $\mu$ that correspond to five fixed values of $\xi$  between $100$ and $200$ nm [dots in Fig.~\ref{fig:spin_splitting}(a)].
This allows us to compare energy spectra computed at different values of $x$, $W$, and $L_N$.
The suppression of $\Delta E$ is a consequence of the reduction of the SOI, as confirmed by Fig. 3(b) in the SM~\cite{Note1}.
At the same time, $\Delta E$ increases with $\xi$, because larger values of $\xi$ correspond to a larger SOI, see Fig.~\ref{fig:spin_splitting}(a).

At $x=0$, $\Delta E$ easily reaches $1~$GHz, comparable to ASQs in InAs nanowires~\cite{tosi2019,hays2020,hays2021,bargerbos2023,Lu2025,PhysRevResearch.3.013036}, while even relatively low values of strain are detrimental for the design of ASQs.
In particular, at $x\gtrsim 10\%$, $\Delta E$ remain smaller than $25$~MHz for all values of $\mu$.
This is consistent with current experiments in $\varepsilon$-Ge with $x=20\%$ that have not observed spin-spitting~\cite{tenkate2025,hinderling2024}.
Interestingly, $\Delta E$ vanishes at $x\approx 4.5\%$ independent of $\xi$. At this concentration, $\delta v$ vanishes at $E_z=1$~mV/nm, see Fig. 3(b) in~\cite{Note1}.

Fig.~\ref{fig:spin_splitting}(c) shows the dependence of $\Delta E$ on the width and length of the junction in unstrained Ge.
The spin-splitting increases with $\xi$ and rapidly saturates with $W$, especially for small values of $\xi$.
The peak at $\xi=175,\, 200$ nm, originates from the enhancement of the Fermi-velocity difference $\delta v$ at the anti-crossing between the ground-state HH and the first LH excited state, as shown in Fig.~\ref{fig:fig1}(b).
The large increase of SOI for squeezed Ge channels is consistent with previous observations~\cite{PhysRevB.104.115425,PhysRevB.105.075308,PhysRevB.106.235408,hoffman2025ge,PhysRevB.110.155431}.
The dependence of $\Delta E$ on the junction length $L_N$ illustrates a cross-over between the short ($L_N \ll \xi$) and long ($L_N\gg \xi$) junction limits. In the former, $\Delta E$ increases with $L_N$, while in the latter it decreases due to crowding of Andreev levels~\cite{nazarov2009}.

Fig.~\ref{fig:spin_splitting}(c) shows that the junction design  is crucial to optimize $\Delta E$.
Importantly, the saturation of $\Delta E$ at large $W$ indicates that wide Ge junctions can support ASQs. This behavior originates from the intrinsically strong cubic-in-momentum Rashba SOI of holes in Ge heterostructures~\cite{PhysRevB.107.035435,PhysRevLett.98.097202,terrazos2021}, which produces a sizable Fermi-velocity difference even in wide junctions.
This contrasts with InAs nanowires, where the SOI is linear in momentum and ASQs rely on the hybridization of the lowest sub-band with transverse modes to generate a Fermi-velocity difference between spin states~\cite{park2017}.

All these numerical results are accurately captured by a simple model describing a ballistic mode with linear dispersion between two superconducting interfaces at a distance $L_N$. In this model, the Andreev levels are the solutions to the transcendental equation~\cite{10.1098/rsta.2018.0140}
\begin{equation}\label{eq:transcendent_ABS_equation}
\arccos\frac{E}{\Delta} - \frac{L_NE}{\hbar v} = \frac{\phi}{2}-n\pi\,,
\end{equation}
where $v$ is the velocity of the mode and $n$ is an integer.
Solving this equation with the spin-dependent values of the velocity extracted from the one-dimensional band structure of the confined channel yields Andreev levels in excellent agreement with the full numerical solutions, see the dashed lines in Fig.~\ref{fig:spin_splitting}.

\begin{figure}[t!]
    \centering
    \includegraphics{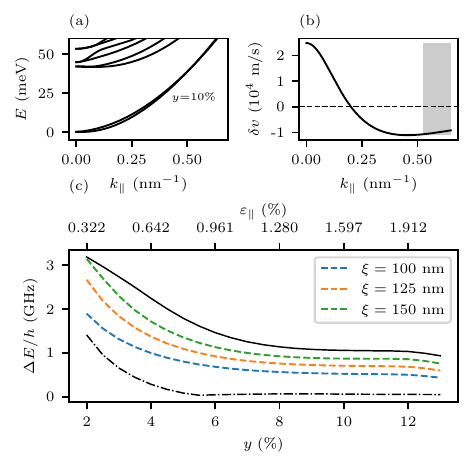}
    \caption{{\emph{Andreev spin splitting in $\bar\varepsilon$-Ge}} (a) Band dispersion of $\bar\varepsilon$-Ge at Sn concentration $y=10\%$, see Fig.~\ref{fig:fig1}, and {{(b)}} corresponding Fermi-velocity difference of the spins of the LH ground-state. {{(c)}} Spin-splitting $\Delta E$ in a $\bar\varepsilon$-Ge junction obtained from Eq.~\eqref{eq:transcendent_ABS_equation} against $y$ for different value of $\xi$. The advantage of $\bar\varepsilon$-Ge is apparent by comparing the black lines representing the maximum $\Delta E$ in the allowed range of $\mu$ between the bottom of the first and second sub-band for $\bar\varepsilon$-Ge (solid line) and unstrained Ge (dash-dotted line). We use $L_N=300$ nm, $W=50$ nm, $\phi=\pi/2$. 
    }
    \label{fig:fig3}
\end{figure}

\paragraph{Andreev spin splitting in $\bar\varepsilon$-Ge. --}
We now consider heterostructures with a Ge$_{1-y}$Sn$_y$ barrier.
The tensile strain favors a LH groundstate that can support a larger SOI than HH~\cite{PhysRevB.110.045409,del2025fully,PhysRevB.107.L161406}.
In analogy to $\varepsilon$-Ge, we derive an effective planar Hamiltonian $H_\mathrm{eff}(\mathbf{k}_\parallel)$ by projecting \eqref{eq.HGe} onto its eigenbasis at $\mathbf{k}_\parallel = \mathbf{0}$.
However, we truncate to the first $20$ HH spin doublets and the first $10$ LH composite spin doublets without including excited levels perturbatively, which prevents the energy dispersion to unphysically bend downwards at large $\textbf{k}_\parallel$.
The resulting Hamiltonian has the same structure of Eq.~\eqref{Heff}, but with $60\times 60$ matrices.
When the Sn concentration $y\gtrsim2\%$, the ground state is a LH (see Fig.~1 in~\cite{Note1}) with dispersion shown in Fig.~\ref{fig:fig3}(a).

Unlike $\varepsilon$-Ge, $\bar\varepsilon$-Ge supports sizeable SOI even at large strain: at $y\approx 10\%$, we find values of $\delta v$ comparable to unstrained Ge [Fig.~\ref{fig:fig3}(b)].
We remark that in contrast to HH in $\varepsilon$-Ge, LH in $\bar\varepsilon$-Ge present an additional sizeable linear-in-momentum SOI, originating from the mixing of the LH and the split-off holes~\cite{PhysRevB.110.045409,PhysRevB.107.L161406}, which yields a finite $\delta v$ at $\textbf{k}_\parallel=\mathbf{0}$.

We estimate the ASQ spin splitting  $\Delta E$ in $\bar\varepsilon$-Ge Josephson junctions using Eq.~\eqref{eq:transcendent_ABS_equation}.
Fig.~\ref{fig:fig3}(c) shows the dependence of $\Delta E$ on $y$ for different $\xi$'s in a junction $W=50$ nm wide.
Comparing the maximal values of $\Delta E$ in $\bar\varepsilon$-Ge (solid black line) with unstrained Ge (dot-dashed black line), we observe that LH in Ge/GeSn outperform HH in Ge/SiGe by a factor of 3 by increasing the maximal spin splitting above $3~$GHz and consistently keeping $\Delta E/h\gtrsim 1$~GHz throughout the available range of $\mu$.

\begin{figure}[t!]
    \centering
    \includegraphics{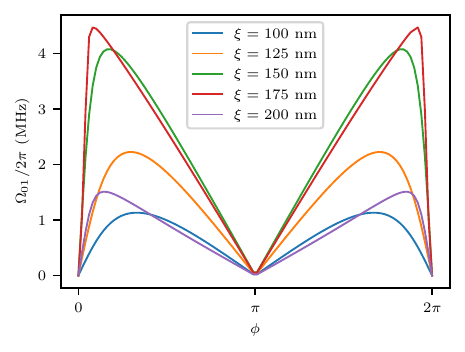}
    \caption{\emph{Driving ASQs.} Rabi frequency $\Omega_{01}$ for the intra-doublet transition $|0\rangle \to |1\rangle$ against $\phi$ for different values of $\xi$. We consider electric-dipole-spin-resonance driving generated by a resonant AC potential with amplitude $\delta V=100\, \mu$V in an unstrained Ge junction with $L_N=300$ nm and $W=100$ nm. For $\xi=175$ nm, this is the same system whose energy levels are shown in Fig.~\ref{fig:fig1}(c).
    }
    \label{fig:fig4}
\end{figure}

\paragraph{Rabi frequency. --} We consider electric-dipole spin-resonance as a source for Rabi oscillations in ASQs~\cite{PhysRevB.74.165319,PhysRevResearch.3.013036}, in which a microwave electric field resonant with $\Delta E$ drives transitions between the qubit states. An alternative, less-efficient driving mechanism relying on the direct coupling of the ASQ to $\phi$ is discussed in the SM~\cite{Note1}.
While LHs are promising candidates for ASQs, the technology for Ge/SiGe heterostructures is currently more mature. We therefore focus on an unstrained Ge channel analogous to the material realized in Ref.~\cite{costa2025} and presented in Figures~\ref{fig:fig1} and~\ref{fig:spin_splitting}.
We consider a small time-dependent transversal electric field  $\delta H(t) = e\delta E_y \cos(\Delta E\,t/\hbar)\,y$ of amplitude $\delta E_y$ aligned to $y$, see Fig.~\ref{fig:fig1}(a), that perturbs the Hamiltonian in Eq.~\eqref{Heff}. 
The gate time $t_g$ is inversely proportional to the Rabi frequency $\Omega_{01}\approx e\delta V\abs{\bra{0} y/W \ket{1}}/\hbar$, where the potential difference across the junction width is $\delta V=\delta E_y W$   and $|\bra{0} y/W \ket{1}|$ is the dimensionless  spin-dipole moment connecting the two ASQ states.
The phase dependence of $\Omega_{01}$ for different values of $\xi$ is shown in Fig.~\ref{fig:fig4}. The Rabi frequency is symmetric around $\phi=\pi$, where $\Omega_{01}=\Delta E=0$ and is larger at small but finite values of $\phi$ where $\Delta E$ is also large, see also Fig.~\ref{fig:fig1}(d). 
The trend against $\xi$ is non-monotonic, in contrast to $\Delta E$, see Fig.~\ref{fig:spin_splitting}, and reaches a maximal around $\xi\sim 175$~nm.
Importantly, realistic values of $\delta V\sim 100~\mu$V yield up to $\Omega_{01}/2\pi\sim 4$~MHz, resulting in fast  $t_g \sim 100$~ns.

\paragraph{Conclusion and outlook. --} Our results emphasize the key role of strain in realizing ASQs in germanium. We find that state-of-the-art  compressively strained $\varepsilon$-Ge does not support large SOI, resulting in limited spin splitting consistent with recent experiments~\cite{tenkate2025,hinderling2024}.  We discuss two alternative platforms, unstrained Ge~\cite{costa2025} and tensile strained $\bar\varepsilon$-Ge~\cite{del2025fully}, where we estimate ASQ energies consistently above 1~GHz, providing viable platforms for high-quality ASQs. 
Further optimization is expected by squeezing the channel~\cite{PhysRevB.104.115425}, engineering inhomogeneous strain~\cite{PhysRevLett.131.097002}, or new material compositions~\cite{delvecchio2026}.
In this respect, a more realistic treatment of confinement potential, charging energy, and superconductor-semiconductor interface will provide a quantitative framework for optimizing Josephson junctions.
Finally, the compatibility of Ge with both superconducting and semiconducting technology in a high-quality and potentially nuclear-spin-free host material raises the prospect that ASQs could be integrated on the same chip with well-developed quantum dot spin qubits~\cite{jakob2025fast}, combining the advantage of both platforms. 

\paragraph{Acknowledgments. --}
We thank members of the Bosco, Scappucci, and Rimbach-Russ groups for valuable discussions. 
VC is grateful for the hospitality of QuTech and TU Delft while carrying out part of the work.
PdV and SB were supported  by the Army Research Office under Award Number: W911NF-23-1-0110 and by NCCR Spin (grant number 225153).
ALRM acknowledges  funding from the European Research Council (Grant
Agreement No.~856526).
VF acknowledges support from the Army Research Office under Grant Number W911NF-22-1-0053. 
The views and conclusions contained in this document are those of the authors and should not be interpreted as representing the official policies, either expressed or implied, of the Army Research Office or the U.S. Government. The U.S. Government is authorized to reproduce and distribute reprints for Government purposes notwithstanding any copyright notation herein.

\paragraph{Data availability. --} The data and code that support the findings of this article are openly available at \url{https://zenodo.org/records/20289119}.

\bibliography{references}

@article{Lu2025,
  title = {{Andreev spin relaxation time in a shadow-evaporated InAs weak link}},
  author = {Lu, Haoran and Bofill, David F. and Sun, Zhenhai and Kanne, Thomas and Nyg\aa{}rd, Jesper and Kjaergaard, Morten and Fatemi, Valla},
  journal = {Phys. Rev. Appl.},
  volume = {24},
  issue = {2},
  pages = {024046},
  numpages = {18},
  year = {2025},
  month = {Aug},
  publisher = {American Physical Society},
  doi = {10.1103/v3lq-t5z8},
  url = {https://link.aps.org/doi/10.1103/v3lq-t5z8}
}

@book{Bir1974,
  title={{Symmetry and Strain-induced Effects in Semiconductors}},
  author={Bir, G.L. and Pikus, G.E.},
  isbn={9780470073216},
  lccn={74014842},
  url={https://books.google.ca/books?id=38m2QgAACAAJ},
  year={1974},
  address={New York},
  publisher={Wiley}
}

@article{Eissfeller2011,
  title = {{Real-space multiband envelope-function approach without spurious solutions}},
  author = {Ei{\ss}feller, T. and Vogl, P.},
  journal = {Phys. Rev. B},
  volume = {84},
  issue = {19},
  pages = {195122},
  numpages = {9},
  year = {2011},
  month = {Nov},
  publisher = {American Physical Society},
  doi = {10.1103/PhysRevB.84.195122},
  url = {https://link.aps.org/doi/10.1103/PhysRevB.84.195122}
}

@book{Winkler2003,
  title={{Spin-orbit Coupling Effects in Two-Dimensional Electron and Hole Systems}},
  author={Winkler, Roland},
  volume={191},
  year={2003},
  publisher={Springer}
}

@article{Bravyi2011,
  title = {{Schrieffer–Wolff transformation for quantum many-body systems}},
  author = {Sergey Bravyi and David P. DiVincenzo and Daniel Loss},
  journal = {Annals of Physics},
  volume = {326},
  number = {10},
  pages = {2793-2826},
  year = {2011},
  issn = {0003-4916},
  doi = {https://doi.org/10.1016/j.aop.2011.06.004},
}

@article{VandeWalle1989,
  title = {{Band lineups and deformation potentials in the model-solid theory}},
  author = {Van de Walle, Chris G.},
  journal = {Phys. Rev. B},
  volume = {39},
  issue = {3},
  pages = {1871--1883},
  numpages = {0},
  year = {1989},
  month = {Jan},
  publisher = {American Physical Society},
  doi = {10.1103/PhysRevB.39.1871},
  url = {https://link.aps.org/doi/10.1103/PhysRevB.39.1871}
}

@article{Reeber1996,
  title = {{Thermal expansion and lattice parameters of group IV semiconductors}},
  journal = "Materials Chemistry and Physics",
  volume = "46",
  number = "2",
  pages = "259 - 264",
  year = "1996",
  issn = "0254-0584",
  doi = "https://doi.org/10.1016/S0254-0584(96)01808-1",
  url = "http://www.sciencedirect.com/science/article/pii/S0254058496018081",
  author = "Robert R. Reeber and Kai Wang",
}

@book{Madelung1991,
  author = {},
  editor = {Otfried Madelung},
  title = {{Semiconductors, Group IV Elements and III-V Compounds}},
  year = {1991},
  publisher = {Springer-Verlag Berlin Heidelberg},
  doi = {10.1007/978-3-642-45681-7},
  issn = {0941-2751}
}

@article{VandeWalle1986,
  title = {{Theoretical calculations of heterojunction discontinuities in the Si/Ge system}},
  author = {Van de Walle, Chris G. and Martin, Richard M.},
  journal = {Phys. Rev. B},
  volume = {34},
  issue = {8},
  pages = {5621--5634},
  numpages = {0},
  year = {1986},
  month = {Oct},
  publisher = {American Physical Society},
  doi = {10.1103/PhysRevB.34.5621},
  url = {https://link.aps.org/doi/10.1103/PhysRevB.34.5621}
}

@article{Lawaetz1971,
  title = {{Valence-Band Parameters in Cubic Semiconductors}},
  author = {Lawaetz, P.},
  journal = {Phys. Rev. B},
  volume = {4},
  issue = {10},
  pages = {3460--3467},
  numpages = {0},
  year = {1971},
  month = {Nov},
  publisher = {American Physical Society},
  doi = {10.1103/PhysRevB.4.3460},
  url = {https://link.aps.org/doi/10.1103/PhysRevB.4.3460}
}

@article{Dismukes1964,
  author={Dismukes, J. P. and Ekstrom, L. and Paff, R. J.},
  title={{Lattice Parameter and Density in Germanium-Silicon Alloys}},
  journal={The Journal of Physical Chemistry},
  year={1964},
  month={Oct},
  day={01},
  publisher={American Chemical Society},
  volume={68},
  number={10},
  pages={3021-3027},
  issn={0022-3654},
  doi={10.1021/j100792a049},
  url={https://doi.org/10.1021/j100792a049}
}

@article{Polak2017,
  doi = {10.1088/1361-6463/aa67bf},
  url = {https://doi.org/10.1088%2F1361-6463%2Faa67bf},
  year = 2017,
  month = {apr},
  publisher = {{IOP} Publishing},
  volume = {50},
  number = {19},
  pages = {195103},
  author = {M P Polak and P Scharoch and R Kudrawiec},
  title = {{The electronic band structure of Ge$_{1-x}$Sn$_x$ in the full composition range: indirect, direct, and inverted gaps regimes, band offsets, and the Burstein{\textendash}Moss effect}},
  journal = {Journal of Physics D: Applied Physics},
}

@article{Menendez2004,
  author = {Men\'endez,J.  and Kouvetakis,J. },
  title = {{Type-I Ge/Ge$_{1-x-y}$Si$_x$Sn$_y$ strained-layer heterostructures with a direct Ge bandgap}},
  journal = {Applied Physics Letters},
  volume = {85},
  number = {7},
  pages = {1175-1177},
  year = {2004},
  doi = {10.1063/1.1784032},
  URL = {https://doi.org/10.1063/1.1784032},
}

@article{Brudevoll1993,
  title = {{Electronic structure of \ensuremath{\alpha}-Sn and its dependence on hydrostatic strain}},
  author = {Brudevoll, T. and Citrin, D. S. and Cardona, M. and Christensen, N. E.},
  journal = {Phys. Rev. B},
  volume = {48},
  issue = {12},
  pages = {8629--8635},
  numpages = {0},
  year = {1993},
  month = {Sep},
  publisher = {American Physical Society},
  doi = {10.1103/PhysRevB.48.8629},
  url = {https://link.aps.org/doi/10.1103/PhysRevB.48.8629}
}

@article{Willatzen1995,
  title = {{Theoretical study of band-edge states in ${\mathrm{Sn}}_{1}$${\mathrm{Ge}}_{\mathit{n}}$ strained-layer superlattices}},
  author = {Willatzen, M. and Lew Yan Voon, L. C. and Santos, P. V. and Cardona, M. and Munzar, D. and Christensen, N. E.},
  journal = {Phys. Rev. B},
  volume = {52},
  issue = {7},
  pages = {5070--5081},
  numpages = {0},
  year = {1995},
  month = {Aug},
  publisher = {American Physical Society},
  doi = {10.1103/PhysRevB.52.5070},
  url = {https://link.aps.org/doi/10.1103/PhysRevB.52.5070}
}

@article{Winkler1996,
  title = {{Theory for the cyclotron resonance of holes in strained asymmetric Ge-SiGe quantum wells}},
  author = {Winkler, R. and Merkler, M. and Darnhofer, T. and R\"ossler, U.},
  journal = {Phys. Rev. B},
  volume = {53},
  issue = {16},
  pages = {10858--10865},
  numpages = {0},
  year = {1996},
  month = {Apr},
  publisher = {American Physical Society},
  doi = {10.1103/PhysRevB.53.10858},
  url = {https://link.aps.org/doi/10.1103/PhysRevB.53.10858}
}

@article{LuLow2012,
  author = {Lu Low,Kain  and Yang,Yue  and Han,Genquan  and Fan,Weijun  and Yeo,Yee-Chia },
  title = {{Electronic band structure and effective mass parameters of Ge1-xSnx alloys}},
  journal = {Journal of Applied Physics},
  volume = {112},
  number = {10},
  pages = {103715},
  year = {2012},
  doi = {10.1063/1.4767381},
  URL = {https://doi.org/10.1063/1.4767381},
}

@article{DelVecchio2024,
  title = {{Light-hole spin confined in germanium}},
  author = {Del Vecchio, Patrick and Moutanabbir, Oussama},
  journal = {Phys. Rev. B},
  volume = {110},
  issue = {4},
  pages = {045409},
  numpages = {11},
  year = {2024},
  month = {Jul},
  publisher = {American Physical Society},
  doi = {10.1103/PhysRevB.110.045409},
  url = {https://link.aps.org/doi/10.1103/PhysRevB.110.045409}
}

@article{chtchelkatchev2003,
  title = {Andreev Quantum Dots for Spin Manipulation},
  author = {Chtchelkatchev, Nikolai M. and Nazarov, Yu. V.},
  journal = {Phys. Rev. Lett.},
  volume = {90},
  issue = {22},
  pages = {226806},
  numpages = {4},
  year = {2003},
  month = {Jun},
  publisher = {American Physical Society},
  doi = {10.1103/PhysRevLett.90.226806},
  url = {https://link.aps.org/doi/10.1103/PhysRevLett.90.226806}
}

@article{beri2006,
  title = {Splitting of {Andreev} levels in a {Josephson} junction by spin-orbit coupling},
  author = {B\'eri, B. and Bardarson, J. H. and Beenakker, C. W. J.},
  journal = {Phys. Rev. B},
  volume = {77},
  issue = {4},
  pages = {045311},
  numpages = {6},
  year = {2008},
  month = {Jan},
  publisher = {American Physical Society},
  doi = {10.1103/PhysRevB.77.045311},
  url = {https://link.aps.org/doi/10.1103/PhysRevB.77.045311}
}

@article{dellanna2008,
  title = {Josephson current through a quantum dot with spin-orbit coupling},
  author = {Dell'Anna, L. and Zazunov, A. and Egger, R. and Martin, T.},
  journal = {Phys. Rev. B},
  volume = {75},
  issue = {8},
  pages = {085305},
  numpages = {7},
  year = {2007},
  month = {Feb},
  publisher = {American Physical Society},
  doi = {10.1103/PhysRevB.75.085305},
  url = {https://link.aps.org/doi/10.1103/PhysRevB.75.085305}
}

@article{michelsen2008,
  title = {Manipulation with {Andreev} states in spin active mesoscopic {Josephson} junctions},
  author = {Michelsen, J. and Shumeiko, V. S. and Wendin, G.},
  journal = {Phys. Rev. B},
  volume = {77},
  issue = {18},
  pages = {184506},
  numpages = {11},
  year = {2008},
  month = {May},
  publisher = {American Physical Society},
  doi = {10.1103/PhysRevB.77.184506},
  url = {https://link.aps.org/doi/10.1103/PhysRevB.77.184506}
}

@article{reynoso2012,
  title = {Spin-orbit-induced chirality of {Andreev} states in {Josephson} junctions},
  author = {Reynoso, Andres A. and Usaj, Gonzalo and Balseiro, C. A. and Feinberg, D. and Avignon, M.},
  journal = {Phys. Rev. B},
  volume = {86},
  issue = {21},
  pages = {214519},
  numpages = {18},
  year = {2012},
  month = {Dec},
  publisher = {American Physical Society},
  doi = {10.1103/PhysRevB.86.214519},
  url = {https://link.aps.org/doi/10.1103/PhysRevB.86.214519}
}

@article{padurariu2010,
  title = {Theoretical proposal for superconducting spin qubits},
  author = {Padurariu, C. and Nazarov, Yu. V.},
  journal = {Phys. Rev. B},
  volume = {81},
  issue = {14},
  pages = {144519},
  numpages = {12},
  year = {2010},
  month = {Apr},
  publisher = {American Physical Society},
  doi = {10.1103/PhysRevB.81.144519},
  url = {https://link.aps.org/doi/10.1103/PhysRevB.81.144519}
}

@article{park2017,
  title = {Andreev spin qubits in multichannel {Rashba} nanowires},
  author = {Park, Sunghun and Yeyati, A. Levy},
  journal = {Phys. Rev. B},
  volume = {96},
  issue = {12},
  pages = {125416},
  numpages = {13},
  year = {2017},
  month = {Sep},
  publisher = {American Physical Society},
  doi = {10.1103/PhysRevB.96.125416},
  url = {https://link.aps.org/doi/10.1103/PhysRevB.96.125416}
}

@article{tosi2019,
  title = {Spin-Orbit Splitting of {Andreev} States Revealed by Microwave Spectroscopy},
  author = {Tosi, L. and Metzger, C. and Goffman, M. F. and Urbina, C. and Pothier, H. and Park, Sunghun and Yeyati, A. Levy and Nyg\aa{}rd, J. and Krogstrup, P.},
  journal = {Phys. Rev. X},
  volume = {9},
  issue = {1},
  pages = {011010},
  numpages = {13},
  year = {2019},
  month = {Jan},
  publisher = {American Physical Society},
  doi = {10.1103/PhysRevX.9.011010},
  url = {https://link.aps.org/doi/10.1103/PhysRevX.9.011010}
}

@article{hays2021,
    author = {M. Hays  and V. Fatemi  and D. Bouman  and J. Cerrillo  and S. Diamond  and K. Serniak  and T. Connolly  and P. Krogstrup  and J. Nygård  and A. Levy Yeyati  and A. Geresdi  and M. H. Devoret },
    title = {Coherent manipulation of an {Andreev} spin qubit},
    journal = {Science},
    volume = {373},
    number = {6553},
    pages = {430-433},
    year = {2021},
    doi = {10.1126/science.abf0345},
    URL = {https://www.science.org/doi/abs/10.1126/science.abf0345}
}

@article{pita2023,
  title={Direct manipulation of a superconducting spin qubit strongly coupled to a transmon qubit},
  author={Pita-Vidal, Marta and Bargerbos, Arno and {\v{Z}}itko, Rok and Splitthoff, Lukas J and Gr{\"u}nhaupt, Lukas and Wesdorp, Jaap J and Liu, Yu and Kouwenhoven, Leo P and Aguado, Ram{\'o}n and van Heck, Bernard and others},
  journal={Nature Physics},
  volume={19},
  number={8},
  pages={1110--1115},
  year={2023},
  publisher={Nature Publishing Group UK London},
  doi={https://doi.org/10.1038/s41567-023-02071-x}
}

@article{pita2024,
  title={Strong tunable coupling between two distant superconducting spin qubits},
  author={Pita-Vidal, Marta and Wesdorp, Jaap J and Splitthoff, Lukas J and Bargerbos, Arno and Liu, Yu and Kouwenhoven, Leo P and Andersen, Christian Kraglund},
  journal={Nature Physics},
  pages={1--6},
  year={2024},
  publisher={Nature Publishing Group UK London},
  doi={https://doi.org/10.1038/s41567-024-02497-x}
}

@article{bargerbos2023,
  title = {Spectroscopy of Spin-Split {Andreev} Levels in a Quantum Dot with Superconducting Leads},
  author = {Bargerbos, Arno and Pita-Vidal, Marta and \ifmmode \check{Z}\else \v{Z}\fi{}itko, Rok and Splitthoff, Lukas J. and Gr\"unhaupt, Lukas and Wesdorp, Jaap J. and Liu, Yu and Kouwenhoven, Leo P. and Aguado, Ram\'on and Andersen, Christian Kraglund and Kou, Angela and van Heck, Bernard},
  journal = {Phys. Rev. Lett.},
  volume = {131},
  issue = {9},
  pages = {097001},
  numpages = {7},
  year = {2023},
  month = {Aug},
  publisher = {American Physical Society},
  doi = {10.1103/PhysRevLett.131.097001},
  url = {https://link.aps.org/doi/10.1103/PhysRevLett.131.097001}
}

@article{hays2020,
  title={Continuous monitoring of a trapped superconducting spin},
  author={Hays, M and Fatemi, V and Serniak, K and Bouman, D and Diamond, S and de Lange, G and Krogstrup, P and Nyg{\aa}rd, J and Geresdi, A and Devoret, MH},
  journal={Nature Physics},
  volume={16},
  number={11},
  pages={1103--1107},
  year={2020},
  doi={https://doi.org/10.1038/s41567-020-0952-3},
  publisher={Nature Publishing Group UK London}
}

@article{scappucci2021,
  title={The germanium quantum information route},
  author={Scappucci, Giordano and Kloeffel, Christoph and Zwanenburg, Floris A and Loss, Daniel and Myronov, Maksym and Zhang, Jian-Jun and De Franceschi, Silvano and Katsaros, Georgios and Veldhorst, Menno},
  journal={Nature Reviews Materials},
  volume={6},
  number={10},
  pages={926--943},
  year={2021},
  doi={https://doi.org/10.1038/s41578-020-00262-z},
  publisher={Nature Publishing Group UK London}
}

@misc{hoffman2025ge,
      title={Resolving {Andreev spin qubits} in germanium-based {Josephson} junctions}, 
      author={Silas Hoffman and Charles Tahan},
      year={2025},
      eprint={2506.13988},
      archivePrefix={arXiv},
      primaryClass={cond-mat.mes-hall},
      url={https://arxiv.org/abs/2506.13988}, 
}

@article{hinderling2024,
  title = {Direct Microwave Spectroscopy of {Andreev} Bound States in Planar {Ge} {Josephson} Junctions},
  author = {Hinderling, M. and ten Kate, S. C. and Coraiola, M. and Haxell, D.Z. and Stiefel, M. and Mergenthaler, M. and Paredes, S. and Bedell, S.W. and Sabonis, D. and Nichele, F.},
  journal = {PRX Quantum},
  volume = {5},
  issue = {3},
  pages = {030357},
  numpages = {14},
  year = {2024},
  month = {Sep},
  publisher = {American Physical Society},
  doi = {10.1103/PRXQuantum.5.030357},
  url = {https://link.aps.org/doi/10.1103/PRXQuantum.5.030357}
}

@article{tenkate2025,
  title = {Finite-length effects and {Coulomb} interaction in {Josephson} junctions based on {Ge} quantum wells and probed with microwave spectroscopy},
  author = {ten Kate, S.C. and Ohnmacht, D.C. and Coraiola, M. and Antonelli, T. and Paredes, S. and Schupp, F.J. and Hinderling, M. and Bedell, S.W. and Belzig, W. and Cuevas, J.C. and Svetogorov, A.E. and Nichele, F. and Sabonis, D.},
  journal = {Phys. Rev. Appl.},
  volume = {24},
  issue = {6},
  pages = {064005},
  numpages = {14},
  year = {2025},
  month = {Dec},
  publisher = {American Physical Society},
  doi = {10.1103/hb1h-8jn9},
  url = {https://link.aps.org/doi/10.1103/hb1h-8jn9}
}

@misc{costa2025,
      title={Buried unstrained germanium channels: a lattice-matched platform for quantum technology}, 
      author={Davide Costa and Patrick Del Vecchio and Karina Hudson and Lucas E. A. Stehouwer and Alberto Tosato and Davide Degli Esposti and Mario Lodari and Stefano Bosco and Giordano Scappucci},
      year={2025},
      eprint={2506.04724},
      archivePrefix={arXiv},
      primaryClass={cond-mat.mes-hall},
      url={https://arxiv.org/abs/2506.04724}, 
}

@article{burkard2023,
  title = {Semiconductor spin qubits},
  author = {Burkard, Guido and Ladd, Thaddeus D. and Pan, Andrew and Nichol, John M. and Petta, Jason R.},
  journal = {Rev. Mod. Phys.},
  volume = {95},
  issue = {2},
  pages = {025003},
  numpages = {58},
  year = {2023},
  month = {Jun},
  publisher = {American Physical Society},
  doi = {10.1103/RevModPhys.95.025003},
  url = {https://link.aps.org/doi/10.1103/RevModPhys.95.025003}
}

@article{blais2021,
  title = {Circuit quantum electrodynamics},
  author = {Blais, Alexandre and Grimsmo, Arne L. and Girvin, S. M. and Wallraff, Andreas},
  journal = {Rev. Mod. Phys.},
  volume = {93},
  issue = {2},
  pages = {025005},
  numpages = {72},
  year = {2021},
  month = {May},
  publisher = {American Physical Society},
  doi = {10.1103/RevModPhys.93.025005},
  url = {https://link.aps.org/doi/10.1103/RevModPhys.93.025005}
}

@article{stano2022,
  title={Review of performance metrics of spin qubits in gated semiconducting nanostructures},
  author={Stano, Peter and Loss, Daniel},
  journal={Nature Reviews Physics},
  volume={4},
  number={10},
  pages={672--688},
  year={2022},
  publisher={Nature Publishing Group UK London},
  doi={https://doi.org/10.1038/s42254-022-00484-w}
}

@article{kwant,
doi = {10.1088/1367-2630/16/6/063065},
url = {https://doi.org/10.1088/1367-2630/16/6/063065},
year = {2014},
month = {jun},
publisher = {IOP Publishing},
volume = {16},
number = {6},
pages = {063065},
author = {Groth, Christoph W and Wimmer, Michael and Akhmerov, Anton R and Waintal, Xavier},
title = {Kwant: a software package for quantum transport},
journal = {New Journal of Physics}
}

@article{PhysRevResearch.3.013036,
  title = {Circuit-QED with phase-biased Josephson weak links},
  author = {Metzger, C. and Park, Sunghun and Tosi, L. and Janvier, C. and Reynoso, A. A. and Goffman, M. F. and Urbina, C. and Levy Yeyati, A. and Pothier, H.},
  journal = {Phys. Rev. Res.},
  volume = {3},
  issue = {1},
  pages = {013036},
  numpages = {20},
  year = {2021},
  month = {Jan},
  publisher = {American Physical Society},
  doi = {10.1103/PhysRevResearch.3.013036},
  url = {https://link.aps.org/doi/10.1103/PhysRevResearch.3.013036}
}

@article{terrazos2021,
  title = {Theory of hole-spin qubits in strained germanium quantum dots},
  author = {Terrazos, L. A. and Marcellina, E. and Wang, Zhanning and Coppersmith, S. N. and Friesen, Mark and Hamilton, A. R. and Hu, Xuedong and Koiller, Belita and Saraiva, A. L. and Culcer, Dimitrie and Capaz, Rodrigo B.},
  journal = {Phys. Rev. B},
  volume = {103},
  issue = {12},
  pages = {125201},
  numpages = {10},
  year = {2021},
  month = {Mar},
  publisher = {American Physical Society},
  doi = {10.1103/PhysRevB.103.125201},
  url = {https://link.aps.org/doi/10.1103/PhysRevB.103.125201}
}

@article{PhysRevLett.98.097202,
  title = {Electric Dipole Spin Resonance for Heavy Holes in Quantum Dots},
  author = {Bulaev, Denis V. and Loss, Daniel},
  journal = {Phys. Rev. Lett.},
  volume = {98},
  issue = {9},
  pages = {097202},
  numpages = {4},
  year = {2007},
  month = {Feb},
  publisher = {American Physical Society},
  doi = {10.1103/PhysRevLett.98.097202},
  url = {https://link.aps.org/doi/10.1103/PhysRevLett.98.097202}
}

@article{PhysRevB.107.035435,
  title = {Planar Josephson junctions in germanium: Effect of cubic spin-orbit interaction},
  author = {Luethi, Melina and Laubscher, Katharina and Bosco, Stefano and Loss, Daniel and Klinovaja, Jelena},
  journal = {Phys. Rev. B},
  volume = {107},
  issue = {3},
  pages = {035435},
  numpages = {19},
  year = {2023},
  month = {Jan},
  publisher = {American Physical Society},
  doi = {10.1103/PhysRevB.107.035435},
  url = {https://link.aps.org/doi/10.1103/PhysRevB.107.035435}
}

@article{10.1098/rsta.2018.0140,
    author = {Sauls, J. A.},
    title = {Andreev bound states and their signatures},
    journal = {Philosophical Transactions of the Royal Society A: Mathematical, Physical and Engineering Sciences},
    volume = {376},
    number = {2125},
    pages = {20180140},
    year = {2018},
    month = {06},
    issn = {1364-503X},
    doi = {10.1098/rsta.2018.0140},
    url = {https://doi.org/10.1098/rsta.2018.0140}
}

@article{PhysRevB.74.165319,
  title = {Electric-dipole-induced spin resonance in quantum dots},
  author = {Golovach, Vitaly N. and Borhani, Massoud and Loss, Daniel},
  journal = {Phys. Rev. B},
  volume = {74},
  issue = {16},
  pages = {165319},
  numpages = {10},
  year = {2006},
  month = {Oct},
  publisher = {American Physical Society},
  doi = {10.1103/PhysRevB.74.165319},
  url = {https://link.aps.org/doi/10.1103/PhysRevB.74.165319}
}

@Article{Kaul2025,
author={Kaul, Prateek
and Karthein, Jan
and Buchhorn, Jonas
and Kawano, Taizo
and Usubuchi, Taisei
and Ishihara, Jun
and Rotaru, Nicolas
and Del Vecchio, Patrick
and Concepcion, Omar
and Ikonic, Zoran
and Gr{\"u}tzmacher, Detlev
and Zhao, Qing-Tai
and Moutanabbir, Oussama
and Kohda, Makoto
and Sch{\"a}pers, Thomas
and Buca, Dan},
title={GeSn quantum wells as a platform for spin-resolved hole transport},
journal={Communications Materials},
year={2025},
month={Oct},
day={02},
volume={6},
number={1},
pages={216},
issn={2662-4443},
doi={10.1038/s43246-025-00934-9},
url={https://doi.org/10.1038/s43246-025-00934-9}
}

@article{PhysRevB.104.115425,
  title = {Squeezed hole spin qubits in Ge quantum dots with ultrafast gates at low power},
  author = {Bosco, Stefano and Benito, M\'onica and Adelsberger, Christoph and Loss, Daniel},
  journal = {Phys. Rev. B},
  volume = {104},
  issue = {11},
  pages = {115425},
  numpages = {6},
  year = {2021},
  month = {Sep},
  publisher = {American Physical Society},
  doi = {10.1103/PhysRevB.104.115425},
  url = {https://link.aps.org/doi/10.1103/PhysRevB.104.115425}
}

@article{PhysRevB.105.075308,
  title = {Hole-spin qubits in Ge nanowire quantum dots: Interplay of orbital magnetic field, strain, and growth direction},
  author = {Adelsberger, Christoph and Benito, M\'onica and Bosco, Stefano and Klinovaja, Jelena and Loss, Daniel},
  journal = {Phys. Rev. B},
  volume = {105},
  issue = {7},
  pages = {075308},
  numpages = {23},
  year = {2022},
  month = {Feb},
  publisher = {American Physical Society},
  doi = {10.1103/PhysRevB.105.075308},
  url = {https://link.aps.org/doi/10.1103/PhysRevB.105.075308}
}

@article{PhysRevB.106.235408,
  title = {Enhanced orbital magnetic field effects in Ge hole nanowires},
  author = {Adelsberger, Christoph and Bosco, Stefano and Klinovaja, Jelena and Loss, Daniel},
  journal = {Phys. Rev. B},
  volume = {106},
  issue = {23},
  pages = {235408},
  numpages = {27},
  year = {2022},
  month = {Dec},
  publisher = {American Physical Society},
  doi = {10.1103/PhysRevB.106.235408},
  url = {https://link.aps.org/doi/10.1103/PhysRevB.106.235408}
}

@article{PhysRevB.110.155431,
  title = {Germanium-based hybrid semiconductor-superconductor topological quantum computing platforms: Disorder effects},
  author = {Laubscher, Katharina and Sau, Jay D. and Das Sarma, Sankar},
  journal = {Phys. Rev. B},
  volume = {110},
  issue = {15},
  pages = {155431},
  numpages = {21},
  year = {2024},
  month = {Oct},
  publisher = {American Physical Society},
  doi = {10.1103/PhysRevB.110.155431},
  url = {https://link.aps.org/doi/10.1103/PhysRevB.110.155431}
}

@article{PhysRevLett.131.097002,
  title = {Hole-Spin Driving by Strain-Induced Spin-Orbit Interactions},
  author = {Abadillo-Uriel, Jos\'e Carlos and Rodr\'{\i}guez-Mena, Esteban A. and Martinez, Biel and Niquet, Yann-Michel},
  journal = {Phys. Rev. Lett.},
  volume = {131},
  issue = {9},
  pages = {097002},
  numpages = {7},
  year = {2023},
  month = {Sep},
  publisher = {American Physical Society},
  doi = {10.1103/PhysRevLett.131.097002},
  url = {https://link.aps.org/doi/10.1103/PhysRevLett.131.097002}
}

@article{delvecchio2026,
  title = {Tailoring Germanium Heterostructures for Quantum Devices with Machine Learning},
  author = {Del Vecchio, P. and Rossi, K and Scappucci, G and Bosco, S },
  year={2026},
  eprint={ },
  journal={arXiv:2604.21732},
  url={https://arxiv.org/abs/2604.21732}
}

@article{jakob2025fast,
  title={Fast readout of quantum dot spin qubits via {Andreev} spins},
  author={Jakob, Mich{\`e}le and Laubscher, Katharina and Del Vecchio, Patrick and Chatterjee, Anasua and Fatemi, Valla and Bosco, Stefano},
  journal={arXiv:2506.19762},
  year={2025},
  url={https://arxiv.org/abs/2506.19762}
}

@book{nazarov2009,
  title={Quantum transport: introduction to nanoscience},
  author={Nazarov, Yuli V and Blanter, Yaroslav M},
  year={2009},
  publisher={Cambridge university press}
}

@article{hoffman2025,
  title = {Decoherence in {Andreev} spin qubits},
  author = {Hoffman, Silas and Hays, Max and Serniak, Kyle and Hazard, Thomas and Tahan, Charles},
  journal = {Phys. Rev. B},
  volume = {111},
  issue = {4},
  pages = {045304},
  numpages = {12},
  year = {2025},
  month = {Jan},
  publisher = {American Physical Society},
  doi = {10.1103/PhysRevB.111.045304},
  url = {https://link.aps.org/doi/10.1103/PhysRevB.111.045304}
}

@article{del2025fully,
  title={Fully tunable strong spin-orbit interactions in light hole germanium quantum channels},
  author={Del Vecchio, Patrick and Bosco, Stefano and Loss, Daniel and Moutanabbir, Oussama},
  journal={arXiv:2506.14759},
  year={2025},
  url={https://arxiv.org/abs/2506.14759}
}

@article{PhysRevB.107.L161406,
  title = {Light-hole gate-defined spin-orbit qubit},
  author = {Del Vecchio, Patrick and Moutanabbir, Oussama},
  journal = {Phys. Rev. B},
  volume = {107},
  issue = {16},
  pages = {L161406},
  numpages = {6},
  year = {2023},
  month = {Apr},
  publisher = {American Physical Society},
  doi = {10.1103/PhysRevB.107.L161406},
  url = {https://link.aps.org/doi/10.1103/PhysRevB.107.L161406}
}

@article{PhysRevB.110.045409,
  title = {Light-hole spin confined in germanium},
  author = {Del Vecchio, Patrick and Moutanabbir, Oussama},
  journal = {Phys. Rev. B},
  volume = {110},
  issue = {4},
  pages = {045409},
  numpages = {11},
  year = {2024},
  month = {Jul},
  publisher = {American Physical Society},
  doi = {10.1103/PhysRevB.110.045409},
  url = {https://link.aps.org/doi/10.1103/PhysRevB.110.045409}
}

@Article{Mauro2025,
author={Mauro, Lorenzo
and Rodr{\'i}guez, Mauricio J.
and Rodr{\'i}guez-Mena, Esteban A.
and Niquet, Yann-Michel},
title={Hole spin qubits in unstrained Germanium layers},
journal={npj Quantum Information},
year={2025},
month={Oct},
day={22},
volume={11},
number={1},
pages={167},
issn={2056-6387},
doi={10.1038/s41534-025-01108-8},
url={https://doi.org/10.1038/s41534-025-01108-8}
}

@article{fabris2026granular,
  title={Granular aluminum induced superconductivity in germanium for hole spin-based hybrid devices},
  author={Fabris, Giorgio and Falthansl-Scheinecker, Paul and Shah, Devashish and Pino, Daniel Michel and Borovkov, Maksim and Bubis, Anton and Roux, Kevin and Sokolova, Dina and Juanes, Alejandro Andres and Costanzo, Tommaso and others},
  journal={arXiv:2602.21364},
  year={2026},
  url={https://arxiv.org/abs/2602.21364}
}

@Article{Lakic2025,
author={Lakic, Lazar
and Lawrie, William I. L.
and van Driel, David
and Stehouwer, Lucas E. A.
and Su, Yao
and Veldhorst, Menno
and Scappucci, Giordano
and Kuemmeth, Ferdinand
and Chatterjee, Anasua},
title={A quantum dot in germanium proximitized by a superconductor},
journal={Nature Materials},
year={2025},
month={Apr},
day={01},
volume={24},
number={4},
pages={552-558},
issn={1476-4660},
doi={10.1038/s41563-024-02095-5},
url={https://doi.org/10.1038/s41563-024-02095-5}
}

@Article{Tosato2023,
author={Tosato, Alberto
and Levajac, Vukan
and Wang, Ji-Yin
and Boor, Casper J.
and Borsoi, Francesco
and Botifoll, Marc
and Borja, Carla N.
and Mart{\'i}-S{\'a}nchez, Sara
and Arbiol, Jordi
and Sammak, Amir
and Veldhorst, Menno
and Scappucci, Giordano},
title={Hard superconducting gap in germanium},
journal={Communications Materials},
year={2023},
month={Apr},
day={06},
volume={4},
number={1},
pages={23},
issn={2662-4443},
doi={10.1038/s43246-023-00351-w},
url={https://doi.org/10.1038/s43246-023-00351-w}
}

@article{mmsd-wfnf,
  title = {Theory of superconducting proximity effect in hole-based hybrid semiconductor-superconductor devices},
  author = {Pino, D. Michel and Souto, Rub\'en Seoane and Calder\'on, Maria Jos\'e and Aguado, Ram\'on and Abadillo-Uriel, Jos\'e Carlos},
  journal = {Phys. Rev. B},
  volume = {111},
  issue = {23},
  pages = {235443},
  numpages = {20},
  year = {2025},
  month = {Jun},
  publisher = {American Physical Society},
  doi = {10.1103/mmsd-wfnf},
  url = {https://link.aps.org/doi/10.1103/mmsd-wfnf}
}

@article{k4jh-pnxy,
  title = {Superconducting proximity effect in two-dimensional hole gases},
  author = {Babkin, Serafim S. and Joecker, Benjamin and Flensberg, Karsten and Serbyn, Maksym and Danon, Jeroen},
  journal = {Phys. Rev. B},
  volume = {111},
  issue = {21},
  pages = {214518},
  numpages = {21},
  year = {2025},
  month = {Jun},
  publisher = {American Physical Society},
  doi = {10.1103/k4jh-pnxy},
  url = {https://link.aps.org/doi/10.1103/k4jh-pnxy}
}

@article{PhysRevB.89.184507,
  title = {Exporting superconductivity across the gap: Proximity effect for semiconductor valence-band states due to contact with a simple-metal superconductor},
  author = {Moghaddam, A. G. and Kernreiter, T. and Governale, M. and Z\"ulicke, U.},
  journal = {Phys. Rev. B},
  volume = {89},
  issue = {18},
  pages = {184507},
  numpages = {9},
  year = {2014},
  month = {May},
  publisher = {American Physical Society},
  doi = {10.1103/PhysRevB.89.184507},
  url = {https://link.aps.org/doi/10.1103/PhysRevB.89.184507}
}

@article{dqgc-8crs,
  title = {Atypical Josephson effect in hybrid superconductor-hole systems},
  author = {Johannsen, Peter D. and Legg, Henry F. and Bosco, Stefano and Loss, Daniel and Klinovaja, Jelena},
  journal = {Phys. Rev. Res.},
  volume = {8},
  issue = {1},
  pages = {013289},
  numpages = {18},
  year = {2026},
  month = {Mar},
  publisher = {American Physical Society},
  doi = {10.1103/dqgc-8crs},
  url = {https://link.aps.org/doi/10.1103/dqgc-8crs}
}

@article{PhysRevB.108.155433,
  title = {Microscopic analysis of proximity-induced superconductivity and metallization effects in superconductor-germanium hole nanowires},
  author = {Adelsberger, Christoph and Legg, Henry F. and Loss, Daniel and Klinovaja, Jelena},
  journal = {Phys. Rev. B},
  volume = {108},
  issue = {15},
  pages = {155433},
  numpages = {13},
  year = {2023},
  month = {Oct},
  publisher = {American Physical Society},
  doi = {10.1103/PhysRevB.108.155433},
  url = {https://link.aps.org/doi/10.1103/PhysRevB.108.155433}
}

@Article{Sagi2024,
author={Sagi, Oliver
and Crippa, Alessandro
and Valentini, Marco
and Janik, Marian
and Baghumyan, Levon
and Fabris, Giorgio
and Kapoor, Lucky
and Hassani, Farid
and Fink, Johannes
and Calcaterra, Stefano
and Chrastina, Daniel
and Isella, Giovanni
and Katsaros, Georgios},
title={A gate tunable transmon qubit in planar Ge},
journal={Nature Communications},
year={2024},
month={Jul},
day={30},
volume={15},
number={1},
pages={6400},
issn={2041-1723},
doi={10.1038/s41467-024-50763-6},
url={https://doi.org/10.1038/s41467-024-50763-6}
}

@article{PhysRevResearch.3.L022005,
  title = {Enhancement of proximity-induced superconductivity in a planar Ge hole gas},
  author = {Aggarwal, Kushagra and Hofmann, Andrea and Jirovec, Daniel and Prieto, Ivan and Sammak, Amir and Botifoll, Marc and Mart\'{\i}-S\'anchez, Sara and Veldhorst, Menno and Arbiol, Jordi and Scappucci, Giordano and Danon, Jeroen and Katsaros, Georgios},
  journal = {Phys. Rev. Res.},
  volume = {3},
  issue = {2},
  pages = {L022005},
  numpages = {7},
  year = {2021},
  month = {Apr},
  publisher = {American Physical Society},
  doi = {10.1103/PhysRevResearch.3.L022005},
  url = {https://link.aps.org/doi/10.1103/PhysRevResearch.3.L022005}
}

@article{kiyooka2026andreev,
  title={Andreev-enhanced conductance quantization and gate-tunable induced superconducting gap in germanium},
  author={Kiyooka, Elyjah and Tangchingchai, Chotivut and Fernandez-Bada, Gonzalo Troncoso and Brun-Barriere, Boris and Zihlmann, Simon and Maurand, Romain and Lefloch, Francois and Schmitt, Vivien and Hartmann, Jean-Michel and Houzet, Manuel and others},
  journal={arXiv:2604.00755},
  year={2026},
  url={https://arxiv.org/pdf/2604.00755}
}

@Article{Kiyooka2025,
author={Kiyooka, Elyjah
and Tangchingchai, Chotivut
and Noirot, Leo
and Leblanc, Axel
and Brun, Boris
and Zihlmann, Simon
and Maurand, Romain
and Schmitt, Vivien
and Dumur, {\'E}tienne
and Hartmann, Jean-Michel
and Lefloch, Francois
and De Franceschi, Silvano},
title={Gatemon Qubit on a Germanium Quantum-Well Heterostructure},
journal={Nano Letters},
year={2025},
month={Jan},
day={08},
publisher={American Chemical Society},
volume={25},
number={1},
pages={562-568},
issn={1530-6984},
doi={10.1021/acs.nanolett.4c05539},
url={https://doi.org/10.1021/acs.nanolett.4c05539}
}

@Article{Leblanc2025,
author={Leblanc, Axel
and Tangchingchai, Chotivut
and Sadre Momtaz, Zahra
and Kiyooka, Elyjah
and Hartmann, Jean-Michel
and Gustavo, Fr{\'e}d{\'e}ric
and Thomassin, Jean-Luc
and Brun, Boris
and Schmitt, Vivien
and Zihlmann, Simon
and Maurand, Romain
and Dumur, {\'E}tienne
and De Franceschi, Silvano
and Lefloch, Fran{\c{c}}ois},
title={Gate- and flux-tunable sin(2$\phi$) Josephson element with planar-Ge junctions},
journal={Nature Communications},
year={2025},
month={Jan},
day={25},
volume={16},
number={1},
pages={1010},
issn={2041-1723},
doi={10.1038/s41467-025-56245-7},
url={https://doi.org/10.1038/s41467-025-56245-7}
}

@Article{Steele2025,
author={Steele, Julian A.
and Strohbeen, Patrick J.
and Verdi, Carla
and Baktash, Ardeshir
and Danilenko, Alisa
and Chen, Yi-Hsun
and van Dijk, Jechiel
and Knudsen, Frederik H.
and Leblanc, Axel
and Perconte, David
and Wang, Lianzhou
and Demler, Eugene
and Salmani-Rezaie, Salva
and Jacobson, Peter
and Shabani, Javad},
title={Superconductivity in substitutional Ga-hyperdoped Ge epitaxial thin films},
journal={Nature Nanotechnology},
year={2025},
month={Dec},
day={01},
volume={20},
number={12},
pages={1757-1763},
issn={1748-3395},
doi={10.1038/s41565-025-02042-8},
url={https://doi.org/10.1038/s41565-025-02042-8}
}

@Article{Hendrickx2018,
author={Hendrickx, N. W.
and Franke, D. P.
and Sammak, A.
and Kouwenhoven, M.
and Sabbagh, D.
and Yeoh, L.
and Li, R.
and Tagliaferri, M. L. V.
and Virgilio, M.
and Capellini, G.
and Scappucci, G.
and Veldhorst, M.},
title={Gate-controlled quantum dots and superconductivity in planar germanium},
journal={Nature Communications},
year={2018},
month={Jul},
day={19},
volume={9},
number={1},
pages={2835},
issn={2041-1723},
doi={10.1038/s41467-018-05299-x},
url={https://doi.org/10.1038/s41467-018-05299-x}
}

@Article{Zhang2025,
author={Zhang, Xin
and Morozova, Elizaveta
and Rimbach-Russ, Maximilian
and Jirovec, Daniel
and Hsiao, Tzu-Kan
and Fari{\~{n}}a, Pablo Cova
and Wang, Chien-An
and Oosterhout, Stefan D.
and Sammak, Amir
and Scappucci, Giordano
and Veldhorst, Menno
and Vandersypen, Lieven M. K.},
title={Universal control of four singlet--triplet qubits},
journal={Nature Nanotechnology},
year={2025},
month={Feb},
day={01},
volume={20},
number={2},
pages={209-215},
issn={1748-3395},
doi={10.1038/s41565-024-01817-9},
url={https://doi.org/10.1038/s41565-024-01817-9}
}

@article{dijkema2026simultaneous,
  title={Simultaneous operation of an 18-qubit modular array in germanium},
  author={Dijkema, Jurgen J and Zhang, Xin and Bardakas, Achilleas and Bouman, Daniel and Cuzzocrea, Alice and van Driel, David and Girardi, Davide and Stehouwer, Lucas EA and Scappucci, Giordano and Zwerver, Anne-Marije J and Hendrickx, Nico W.},
  journal={arXiv:2604.01063},
  year={2026},
  url={https://arxiv.org/abs/2604.01063}
}

@Article{John2025,
author={John, Valentin
and Yu, C{\'e}cile X.
and van Straaten, Barnaby
and Rodr{\'i}guez-Mena, Esteban A.
and Rodr{\'i}guez, Mauricio
and Oosterhout, Stefan D.
and Stehouwer, Lucas E. A.
and Scappucci, Giordano
and Rimbach-Russ, Maximilian
and Bosco, Stefano
and Borsoi, Francesco
and Niquet, Yann-Michel
and Veldhorst, Menno},
title={Robust and localised control of a 10-spin qubit array in germanium},
journal={Nature Communications},
year={2025},
month={Nov},
day={26},
volume={16},
number={1},
pages={10560},
issn={2041-1723},
doi={10.1038/s41467-025-65577-3},
url={https://doi.org/10.1038/s41467-025-65577-3}
}

@Article{Hendrickx2021,
author={Hendrickx, Nico W.
and Lawrie, William I. L.
and Russ, Maximilian
and van Riggelen, Floor
and de Snoo, Sander L.
and Schouten, Raymond N.
and Sammak, Amir
and Scappucci, Giordano
and Veldhorst, Menno},
title={A four-qubit germanium quantum processor},
journal={Nature},
year={2021},
month={Mar},
day={01},
volume={591},
number={7851},
pages={580-585},
issn={1476-4687},
doi={10.1038/s41586-021-03332-6},
url={https://doi.org/10.1038/s41586-021-03332-6}
}

@Article{Borsoi2024,
author={Borsoi, Francesco
and Hendrickx, Nico W.
and John, Valentin
and Meyer, Marcel
and Motz, Sayr
and van Riggelen, Floor
and Sammak, Amir
and de Snoo, Sander L.
and Scappucci, Giordano
and Veldhorst, Menno},
title={Shared control of a 16{\thinspace}semiconductor quantum dot crossbar array},
journal={Nature Nanotechnology},
year={2024},
month={Jan},
day={01},
volume={19},
number={1},
pages={21-27},
issn={1748-3395},
doi={10.1038/s41565-023-01491-3},
url={https://doi.org/10.1038/s41565-023-01491-3}
}

@article{PhysRevB.101.115302,
  title = {First-principles hyperfine tensors for electrons and holes in GaAs and silicon},
  author = {Philippopoulos, Pericles and Chesi, Stefano and Coish, W. A.},
  journal = {Phys. Rev. B},
  volume = {101},
  issue = {11},
  pages = {115302},
  numpages = {15},
  year = {2020},
  month = {Mar},
  publisher = {American Physical Society},
  doi = {10.1103/PhysRevB.101.115302},
  url = {https://link.aps.org/doi/10.1103/PhysRevB.101.115302}
}

@article{PhysRevLett.127.190501,
  title = {Fully Tunable Hyperfine Interactions of Hole Spin Qubits in Si and Ge Quantum Dots},
  author = {Bosco, Stefano and Loss, Daniel},
  journal = {Phys. Rev. Lett.},
  volume = {127},
  issue = {19},
  pages = {190501},
  numpages = {7},
  year = {2021},
  month = {Nov},
  publisher = {American Physical Society},
  doi = {10.1103/PhysRevLett.127.190501},
  url = {https://link.aps.org/doi/10.1103/PhysRevLett.127.190501}
}

@article{PhysRevB.78.155329,
  title = {Spin decoherence of a heavy hole coupled to nuclear spins in a quantum dot},
  author = {Fischer, Jan and Coish, W. A. and Bulaev, D. V. and Loss, Daniel},
  journal = {Phys. Rev. B},
  volume = {78},
  issue = {15},
  pages = {155329},
  numpages = {9},
  year = {2008},
  month = {Oct},
  publisher = {American Physical Society},
  doi = {10.1103/PhysRevB.78.155329},
  url = {https://link.aps.org/doi/10.1103/PhysRevB.78.155329}
}

@article{https://doi.org/10.1002/adma.202305703,
author = {Moutanabbir, Oussama and Assali, Simone and Attiaoui, Anis and Daligou, Gérard and Daoust, Patrick and Vecchio, Patrick Del and Koelling, Sebastian and Luo, Lu and Rotaru, Nicolas},
title = {Nuclear Spin-Depleted, Isotopically Enriched 70Ge/28Si70Ge Quantum Wells},
journal = {Advanced Materials},
volume = {36},
number = {8},
pages = {2305703},
doi = {https://doi.org/10.1002/adma.202305703},
url = {https://advanced.onlinelibrary.wiley.com/doi/abs/10.1002/adma.202305703},
year = {2024}
}

\end{document}


\title{Supplemental Material: \\
Strain engineering of Andreev spin qubits in Germanium}
\author{Vittorio Coppini}
\email{vittorio.coppini@uniroma1.it}
\affiliation{Dipartimento di Fisica, Sapienza Università di Roma, Piazzale Aldo Moro 2, 00185 Rome, Italy}
\author{Patrick Del Vecchio}
\email{p.delvecchio@tudelft.nl}
\affiliation{QuTech and Kavli Institute of Nanoscience, Delft University of Technology, Delft, Netherlands}
\author{Antonio L. R. Manesco}
\affiliation{Kavli Institute of Nanoscience, Delft University of Technology, Delft, Netherlands}
\affiliation{Center for Quantum Devices, Niels Bohr Institute, University of Copenhagen, DK-2100 Copenhagen, Denmark}
\author{Anton Akhmerov}
\affiliation{Kavli Institute of Nanoscience, Delft University of Technology, Delft, Netherlands}
\author{Valla Fatemi}
\affiliation{School of Applied and Engineering Physics, Cornell University, Ithaca, NY, 14853, USA
}
\author{Bernard van Heck}
\affiliation{Dipartimento di Fisica, Sapienza Università di Roma, Piazzale Aldo Moro 2, 00185 Rome, Italy}
\author{Stefano Bosco}
\affiliation{QuTech and Kavli Institute of Nanoscience, Delft University of Technology, Delft, Netherlands}

\begin{abstract}
This Supplemental Material provides a description of the $k\cdot p$ and perturbation theory employed to model the electronic dispersion of the different germanium (Ge) heterostructures. Additional information on the Rashba spin-orbit interaction (SOI) parameters and  Fermi velocities is also provided, as well as details on the material parametrization used in the $k\cdot p$ simulations. We also provide details on the numerical model of the Andreev spin qubit (ASQ) and  the calculation of the matrix elements of $\partial H/\partial \phi$.
\end{abstract}

\maketitle

\section{\texorpdfstring{General 6-band $k\cdot p$ Hamiltonian}{}}\label{sec:kp}
The electronic band structure of the heterostructures are evaluated from $6$-band $k\cdot p$ theory. We consider a two-dimensional hole gas (2DHG) under the influence of a gate electric field $\mathbf{E} = E_z\mathbf{e}_z$ applied along the $z$ direction, which is perpendicular to the 2DHG plane. Here we use the minus sign convention for the hole band structure. To agree with literature, the energy dispersion provided in the main text are reverted in their sign, i.e. holes have a positive effective mass. The details of the two heterostructures used for the numerical calculations are shown in Table \ref{tab:stacks}. In both cases the bottom layer sets the in-plane lattice constant across the stack.

\begin{table}[ht!]
    \centering
    \caption{Ge heterostructures.}
    \begin{threeparttable}
        \begin{tabular}{c c c}
        \toprule
        & $\varepsilon$-Ge (or Ge) & $\bar{\varepsilon}$-Ge \\
        \midrule
         Top layer & Si$_{0.2}$Ge$_{0.8}$ ($10\,\mathrm{nm}$) & Si$_{0.2}$Ge$_{0.8}$ ($10\,\mathrm{nm}$) \\
         Middle layer & Ge ($15\,\mathrm{nm}$) & Ge ($15\,\mathrm{nm}$) \\
         \multirow{2}{*}{Bottom layer (unstrained)} & \SiGe{} ($50\,\mathrm{nm}$) & \GeSn{} ($50\,\mathrm{nm}$) \\
         & $x\in [0, 20]\%$ & $y\in[2,13]\%$
    \end{tabular}
    \label{tab:stacks}
    \end{threeparttable}
\end{table}

The total Hamiltonian $H_\mathrm{Ge}$ can be written as a sum of multiple contributions: (a) a Luttinger-Kohn term $H_\mathrm{LK} = \mathcal{E}_{\Gamma_5^+} + H_k(\mathbf{k}) + H_\mathrm{so}$ that includes the valence band edge energy at the $\Gamma$-point \emph{without} spin-orbit interaction $\mathcal{E}_{\Gamma_5^+}(z)$, a $k$-dependent part $H_k(\mathbf{k})$ and a spin-orbit part $H_\mathrm{so}$; (b) an epitaxial strain part $H_\mathrm{BP}$; and (c) a potential energy part $-eE_zz$. We write $\mathbf{k} = k_x\mathbf{e}_x + k_y\mathbf{e}_y + k_z\mathbf{e}_z$, $\mathbf{k}_\parallel = k_x\mathbf{e}_x + k_y\mathbf{e}_y$ and $k_\pm = k_x \pm ik_y$. We first write the Hamiltonian in the basis of the bulk Bloch states at the $\Gamma$-point in the cartesian representation~\cite{Eissfeller2011}:

\begin{equation}
    \mathcal{B}_X = \left\{\ket{p_x+},\ket{p_y+},\ket{p_z+},\ket{p_x-},\ket{p_y-},\ket{p_z-}\right\}.
\end{equation}

The kinetic term is

\begin{equation}
    H_k(\mathbf{k}) = \mathbb{1}_{2\times 2}\otimes H^k_{vv}(\mathbf{k}),
\end{equation}

\noindent with

\begin{align}
    H_{vv}^k(\mathbf{k}) = \begin{bmatrix}
        k_xLk_x+k_yMk_y+k_zMk_z & k_xN_+k_y+k_yN_-k_x & k_xN_+k_z+k_zN_-k_x \\
        k_yN_+k_x+k_xN_-k_y & k_xMk_x+k_yLk_y+k_zMk_z & k_yN_+k_z+k_zN_-k_y \\
        k_zN_+k_x+k_xN_-k_z & k_zN_+k_y+k_yN_-k_z & k_xMk_x+k_yMk_y+k_zLk_z
    \end{bmatrix},
\end{align}

\noindent where $L$, $M$ and $N_\pm$ are related to the Luttinger parameters $\gamma_{1,2,3}$ and $\kappa$:

\begin{equation}
    \begin{bmatrix}
        L \\
        M \\
        N_+ + \alpha_0 \\
        N_- - \alpha_0
    \end{bmatrix} = -\alpha_0\begin{bmatrix}
        1 & 4 & 0 & 0 \\
        1 & -2 & 0 & 0 \\
        0 & 0 & 3 & 3 \\
        0 & 0 & 3 & -3
    \end{bmatrix}\begin{bmatrix}
        \gamma_1 \\
        \gamma_2 \\
        \gamma_3 \\
        \kappa
    \end{bmatrix},
\end{equation}

\noindent where $\alpha_0 \equiv \hbar^2/(2m_0)$. The spin-orbit term is proportional to the bulk split-off gap $\Delta_0$ and given by

\begin{equation}
    H_\mathrm{so} = \frac{\Delta_0}{3}\begin{bmatrix}
        0 & -i & 0 & 0 & 0 & 1 \\
        i & 0 & 0 & 0 & 0 & -i \\
        0 & 0 & 0 & -1 & i & 0 \\
        0 & 0 & -1 & 0 & i & 0 \\
        0 & 0 & -i & -i & 0 & 0 \\
        1 & i & 0 & 0 & 0 & 0 \\
    \end{bmatrix}.
\end{equation}

\noindent  $H_\mathrm{so}$ has eigenvalues $0$ ($\times 2$), $-2\Delta_0/3$ ($\times 2$) and $+\Delta_0/3$ ($\times 4$). Finally, the strain term is $H_\mathrm{BP} = \mathbb{1}_{2\times 2}\otimes H_{vv}^\varepsilon$, where (assuming a diagonal strain tensor)

\begin{equation}
    H_{vv}^\varepsilon = \begin{bmatrix}
        l\varepsilon_{xx} + m\varepsilon_{yy} + m\varepsilon_{zz} & 0 & 0 \\
        0 & m\varepsilon_{xx} + l\varepsilon_{yy} + m\varepsilon_{zz} & 0 \\
        0 & 0 & m\varepsilon_{xx} + m\varepsilon_{yy} + l\varepsilon_{zz} \\
    \end{bmatrix},
\end{equation}

\noindent where $\varepsilon_{ij}$ are the components of the strain tensor and $l$ and $m$ are related to the $a_v$ and $b$ deformation potential constants:

\begin{equation}
    \begin{bmatrix}
        l \\
        m
    \end{bmatrix} = \begin{bmatrix}
        1 & 2 \\
        1 & -1
    \end{bmatrix}\begin{bmatrix}
        a_v \\
        b
    \end{bmatrix}.
\end{equation}

Note that the ordering between material constants (e.g. $\gamma_1$, $a_v$, $\mathcal{E}_{\Gamma_5^+}$, etc \dots) and the wavevector operator component $k_z$ must be preserved as indicated, since it does not commute with position-dependent material parameters. In fact, if $\gamma\ket{z} = \ket{z}\gamma(z)$ where $\gamma(z)$ is the value of $\gamma$ at position $z$, then for any envelope function $\braket{z|\psi} = \psi(z)$: $\bra{z}[\gamma,k_z]\ket{\psi} = i\gamma^\prime(z)\psi(z)$ where $\gamma^\prime(z) = \frac{d}{dz}\gamma(z)$.

The $k\cdot p$ Hamiltonian is instead often written in a basis that diagonalizes $H_\mathrm{so}$. A  conventional basis is the following~\cite{Winkler2003}:

\begin{equation}
    \mathcal{B}_J = \left\{\Ket{\frac{3}{2},\frac{3}{2}},\Ket{\frac{3}{2},\frac{1}{2}},\Ket{\frac{3}{2},-\frac{1}{2}},\Ket{\frac{3}{2},-\frac{3}{2}},\Ket{\frac{1}{2},\frac{1}{2}},\Ket{\frac{1}{2},-\frac{1}{2}}\right\},
\end{equation}

\noindent where

\begin{align}
    \Ket{\frac{3}{2},\frac{3}{2}}\equiv\Ket{\mathrm{HH}+} &= -\frac{1}{\sqrt{2}}\left(\ket{p_x+}+i\ket{p_y+}\right), \\
    \Ket{\frac{3}{2},\frac{1}{2}}\equiv\Ket{\mathrm{LH}+} &= \frac{1}{\sqrt{6}}\left(2\ket{p_z+}-\ket{p_x-}-i\ket{p_y-}\right), \\
    \Ket{\frac{3}{2},-\frac{1}{2}}\equiv\Ket{\mathrm{LH}-} &= \frac{1}{\sqrt{6}}\left(\ket{p_x+}-i\ket{p_y+}+2\ket{p_z-}\right), \\
    \Ket{\frac{3}{2},-\frac{3}{2}}\equiv\Ket{\mathrm{HH}-} &= \frac{1}{\sqrt{2}}\left(\ket{p_x-}-i\ket{p_y-}\right), \\
    \Ket{\frac{1}{2},\frac{1}{2}}\equiv\Ket{\mathrm{SO}+} &= -\frac{1}{\sqrt{3}}\left(\ket{p_z+}+\ket{p_x-}+i\ket{p_y-}\right), \\
    \Ket{\frac{1}{2},-\frac{1}{2}}\equiv\Ket{\mathrm{SO}-} &= -\frac{1}{\sqrt{3}}\left(\ket{p_x+}-i\ket{p_y+}-\ket{p_z-}\right).
\end{align}

\noindent The unitary $U_{JX}$ connecting the two bases is therefore

\begin{equation}
    U_{JX} = \begin{bmatrix}
        -s_2 & is_2 & 0 & 0 & 0 & 0 \\
        0 & 0 & s_{23} & -s_6 & is_6 & 0 \\
        s_6 & is_6 & 0 & 0 & 0 & s_{23} \\
        0 & 0 & 0 & s_2 & is_2 & 0 \\
        0 & 0 & -s_3 & -s_3 & is_3 & 0 \\
        -s_3 & -is_3 & 0 & 0 & 0 & s_3 \\
    \end{bmatrix},
\end{equation}

\noindent with $s_2=1/\sqrt{2}$, $s_3 = 1/\sqrt{3}$, $s_6 = 1/\sqrt{6}$ and $s_{23} = \sqrt{2/3}$.

\section{\texorpdfstring{First diagonalization of the $k\cdot p$ matrix at $\mathbf{k}_\parallel = 0$}{}}\label{sec:kp_}
We first study the $6$-band $k\cdot p$ matrix at $k_x = k_y = 0$. Its eigenstates provide an orthonormal basis on which the full Hamiltonian at $\mathbf{k}_\parallel \neq 0$ is projected. We neglect cubic corrections to the band structure by assuming $\gamma_2-\gamma_3\approx 0$, i.e. $\gamma_{2,3}\to\tilde{\gamma} = (\gamma_2 + \gamma_3)/2$. In a basis similar to $\mathcal{B}_J$ but with a different ordering:

\begin{equation}
    \mathcal{B}_\mathcal{J} = \left\{\ket{\mathrm{LH}+},\ket{\mathrm{SO}+},\ket{\mathrm{HH}+},\ket{\mathrm{LH}-},\ket{\mathrm{SO}-},\ket{\mathrm{HH}-}\right\},
\end{equation}

\noindent the Hamiltonian $H_0\equiv H_\mathrm{Ge}(k_x=k_y=0)$ takes the following form:

\begin{align}\label{H0}
  H_0 &= \begin{bmatrix}
    H_{\sigma=+} & 0 \\
    0 & H_{\sigma=-}
  \end{bmatrix} \\
  H_\sigma &= H_\sigma^k + H_\sigma^\varepsilon + V_z,
\end{align}

\noindent where

\begin{align}
  H_\sigma^k &= \alpha_0\begin{bmatrix}
  -k_z(\gamma_1+2\tilde{\gamma})k_z & 2\sqrt{2}\sigma k_z\tilde{\gamma}k_z & 0 \\
  & -k_z\gamma_1k_z & 0 \\
  \dagger & & -k_z(\gamma_1-2\tilde{\gamma})k_z
  \end{bmatrix}, \\
  H_\sigma^\varepsilon &= a_v\mathrm{Tr}\,\varepsilon + b\cdot\delta\varepsilon\begin{bmatrix}
  -1 & \sqrt{2}\sigma & 0 \\
  & 0 & 0 \\
  \dagger & & 1
  \end{bmatrix}, \\
  V_z &= \mathcal{E}_{\Gamma_5^+} + \frac{\Delta_0}{3} - eE_zz - \Delta_0\begin{bmatrix}
  0 & 0 & 0 \\
  & 1 & 0 \\
  \dagger & & 0
  \end{bmatrix},
\end{align}

\noindent with $\delta\varepsilon = \varepsilon_{xx} - \varepsilon_{zz}$. We assume that all layers are pseudomorphic to each other, i.e., the in-plane lattice constant does not change along the growth direction and the strain tensor elements satisfy $\varepsilon_{xx}=\varepsilon_{yy}\equiv\varepsilon_\parallel$. We estimate $\varepsilon_\parallel$ in layer $i$ using the following equation:

\begin{equation}
    \varepsilon_\parallel^i = \frac{a_\parallel}{a_0^i} - 1,
\end{equation}

\noindent where $a_\parallel$ is the global in-plane lattice constant set by the bottom layer, and $a_0^i$ is the lattice constant of unstrained material $i$. The out-of-plane component of the strain tensor $\varepsilon_{zz}$ in layer $i$ is evaluated using $\varepsilon_{zz}^i = -(2c_{12}^i/c_{11}^i)\varepsilon_\parallel^i$. The parametrization of the material constants such as the Luttinger parameters and the strain deformation potentials is described in section \ref{sec:mtr}. We stress again that the ordering between $k_z$ and material parameters must be preserved.

\begin{figure}
    \centering
    \includegraphics[width=0.4\linewidth]{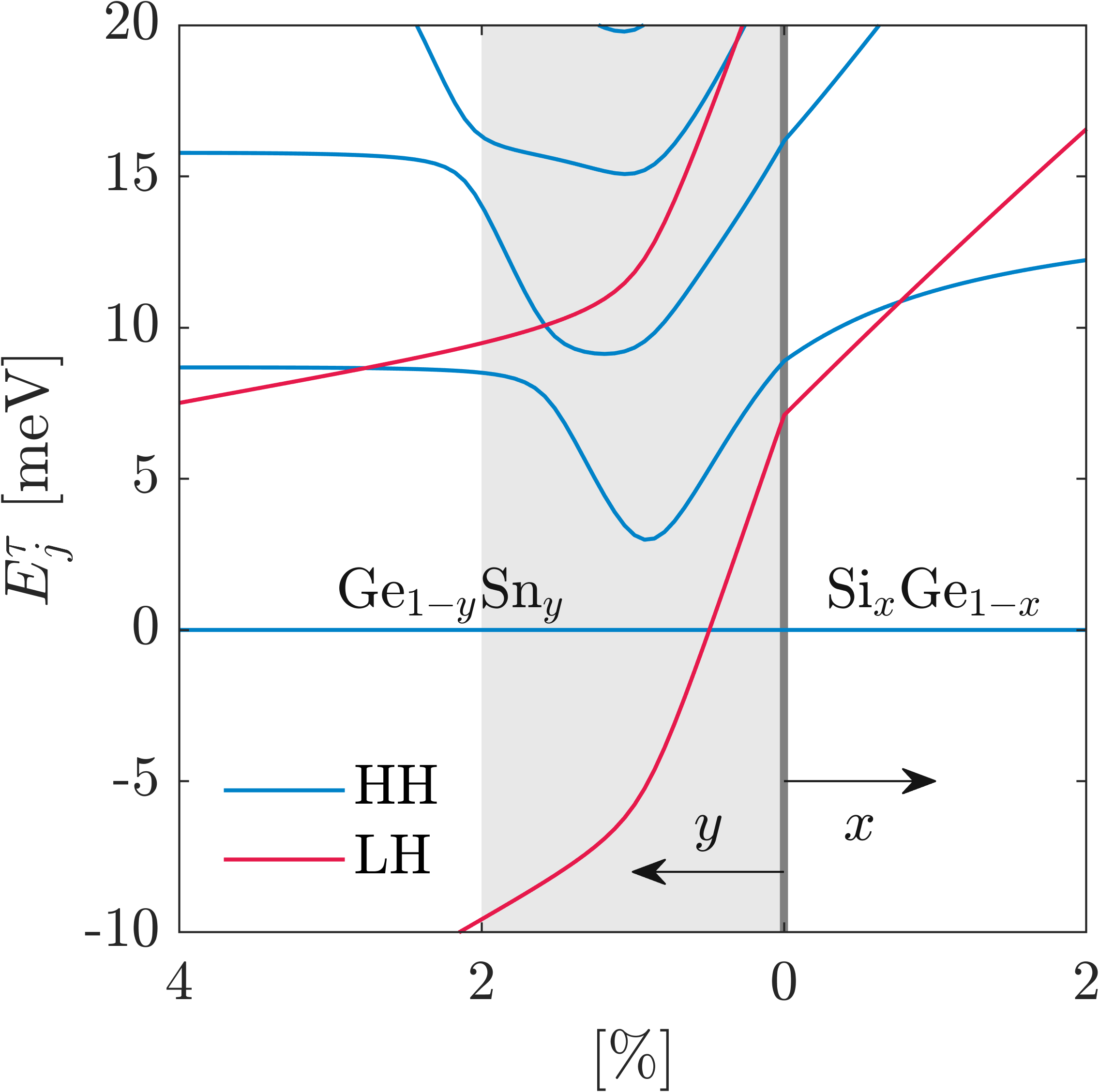}
    \caption{Energies $E_j^\tau$ as a function of the concentration $x$ and $y$. The zero of energy is pinned to the HH ground state, i.e. $E_1^\h{}\equiv0$. The shaded gray area marks the region excluded from the analysis in the main text to avoid the crossing between  HH and  LH ground states.}
    \label{fig:Exy}
\end{figure}

The block-diagonal structure of Eq.~\eqref{H0} suggests that its eigenstates are of two types ($2\times 2 = 4$ including pseudospin $\sigma$):

\begin{subequations}\label{types}
  \begin{align}
    \ket{\h{l};\sigma} &= \Ket{\frac{3}{2}, \frac{3\sigma}{2}}\ket{f^h_l}, \\
    \ket{\e{j};\sigma} &= \Ket{\frac{3}{2}, \frac{\sigma}{2}}\ket{f^\ell_j} + \sigma\Ket{\frac{1}{2}, \frac{\sigma}{2}}\ket{f^s_j},
  \end{align}
\end{subequations}

\noindent where the first ($\h{}$ states) consists of pure HH sub-bands with associated envelope functions $\braket{z|f_l^h}$ and spin-independent energies $E_l^\h{}$, and where the second ($\e{}$ states) consists of superpositions of pure LH and pure split-off holes with envelope functions $\braket{z|f^\ell_j}$ and $\braket{z|f^s_j}$, respectively, with spin-independent energies $E_j^\e{}$. Here, $l$ and $j$ are sub-band indices for $\h{}$ and $\e{}$ sub-bands, respectively. To compute the envelopes and energies, we start from the spin-up ($\sigma=+$) block $H_+$ in Eq.~\eqref{H0}, which we diagonalize for different heterostructure compositions by the finite difference method using a mesh-spacing $\delta z = 0.02\,\mathrm{nm}$. The eigenstates of $H_-$ are the time-reversed conjugates of the eigenstates of $H_+$. Figure \ref{fig:Exy} shows the energies $E_j^\tau$ ($\tau=\{\h{},\e{}\}$) for the first few $\h{}$ and $\e{}$ sub-bands as a function of the Si content $x$ and of the Sn content $y$ near the region where the ground state character changes between HH-like to LH-like at $y\approx0.5\%$. There is a discontinuity in the derivatives $d E_j^\tau/dx|_{x=0}$ and $d E_j^\tau/dy|_{y=0}$ at $x=y=0$ since the rate of change of the lattice constant of \SiGe{} is smaller close to $x=0$ than that of \GeSn{} close to $y=0$. In subsequent calculations and in the main text, we restrict ourselves to the region outside the shaded gray area to avoid the crossing between the HH ground state and the LH ground state, which makes deriving a low-energy effective Hamiltonian more challenging. We extend this buffer until $y=2\%$ because residual strains in the Ge layer can move the transition point to higher Sn concentrations $y$~\cite{DelVecchio2024}.

Using the eigenstates of $H_0$ as the new basis for the Hilbert space, the full 2DHG Hamiltonian $H_\mathrm{Ge}$ for $\mathbf{k}_\parallel\neq 0$ takes the following form (bold symbols indicate that we are in the eigenbasis of $H_0$):

\begin{equation}\label{Hk}
\begin{split}
    \mathbf{H}_\mathrm{Ge}(\mathbf{k}_\parallel) = \mathbf{E}_0 &+ \alpha_0\left[\mathbf{M}_\gamma k_\parallel^2 + \left(i\mathbf{M}_1 k_- + \mathbf{M}_2k_-^2 + \text{h.c.}\right)\right] \\
\end{split}
\end{equation}

\noindent where $k_\parallel^2 = k_x^2 + k_y^2 = k_-k_+$. The Hamiltonian in Eq. \eqref{Hk} implicitly depends on strain and on the SOI strength $\Delta_0$ through the envelopes $\ket{f_j^{\ell,s,h}}$ and the energies $E_j^\tau$. Those terms do not explicitly appear in \eqref{Hk} because the $k\cdot p$ Hamiltonian does not contain terms such as $\Delta_0k_\pm$ or $\delta\varepsilon\cdot k_\pm$. It also integrates out the $z$ dimension through the $\mathbf{M}$-matrices. We have

\begin{equation}
  \mathbf{E}_0 = \begin{bmatrix}
    \mathbf{E}^\h{} & 0 & 0 & 0 \\
    0 & \mathbf{E}^\e{} & 0 & 0 \\
    0 & 0 & \mathbf{E}^\e{} & 0 \\
    0 & 0 & 0 & \mathbf{E}^\h{}
    \end{bmatrix},
\end{equation}

\noindent with $\mathbf{E}^\tau = \mathrm{diag}\{E_1^\tau,E_2^\tau,\dots\}$ ($\tau = \{\h{},\e{}\}$) are the energies of $H_0$. The $\mathbf{M}$-matrices are

\begin{align}
  \mathbf{M}_\gamma &= \begin{bmatrix}
    \boldsymbol{\Gamma}_\parallel^\h{} & 0 & 0 & 0 \\
    0 & \boldsymbol{\Gamma}_\parallel^\e{} & 0 & 0 \\
    0 & 0 & \boldsymbol{\Gamma}_\parallel^\e{} & 0 \\
    0 & 0 & 0 & \boldsymbol{\Gamma}_\parallel^\h{}
  \end{bmatrix}, & \mathbf{M}_1 &= \begin{bmatrix}
    0 & \mathbf{T}^\times & 0 & 0 \\
    0 & 0 & \mathbf{T}^\e{} & 0 \\
    0 & 0 & 0 & \mathbf{T}^{\times\dagger} \\
    0 & 0 & 0 & 0
  \end{bmatrix},
  &\mathbf{M}_2 &= \begin{bmatrix}
    0 & 0 & \boldsymbol{\upmu} & 0 \\
    0 & 0 & 0 & \boldsymbol{\upmu}^\dagger \\
    0 & 0 & 0 & 0 \\
    0 & 0 & 0 & 0
  \end{bmatrix}.
\end{align}

The non-zero blocks of these matrices are all explicitly expanded in terms of the eigenstates of $H_0$:

\begin{subequations}
  \begin{align}
    \Gamma_{\parallel l,l^\prime}^\h{} &= -\braket{f^h_l|\gamma_1 + \tilde{\gamma}|f^h_{l^\prime}}, \\
    \Gamma_{\parallel j,j^\prime}^\e{} &= -\braket{f^\perp_j|\gamma_1-2\tilde{\gamma}|f^\perp_{j^\prime}} - \braket{f^\circ_j|\gamma_1 + \tilde{\gamma}|f^\circ_{j^\prime}}, \\
    T_{l,j}^\times &= -\frac{3i}{\sqrt{2}}\bra{f^h_l}u_+\ket{f^\perp_j}, \\
    T_{j,j^\prime}^\e{} &= -\frac{3i}{\sqrt{2}}\left(\braket{f_j^\circ|u_+|f_{j^\prime}^\perp} - \braket{f_j^\perp|u_-|f_{j^\prime}^\circ}\right), \\
    \mu_{l,j} &= 3\braket{f_l^h|\tilde{\gamma}|f_j^\circ},
  \end{align}
\end{subequations}

\noindent where $u_\pm = \{\tilde{\gamma},k_z\} \pm [\kappa,k_z]$, $\{A,B\} = AB+BA$ and

\begin{align}
  \ket{f_j^\perp} &\equiv \frac{1}{\sqrt{3}}\left(\sqrt{2}\ket{f_j^\ell} - \ket{f_j^s}\right),
  &\ket{f_j^\circ} &\equiv \frac{1}{\sqrt{3}}\left(\ket{f_j^\ell} + \sqrt{2}\ket{f_j^s}\right).
\end{align}

\section{Perturbation theory}\label{sec:pert}
\subsection{Band dispersion}

Away from $\mathbf{k}_\parallel = 0$, Hamiltonian \eqref{Hk} becomes non-diagonal, and folding it down onto any given set of sub-bands, labeled as the $\mathcal{A}$ set, can be performed by means of a Schrieffer-Wolff transformation (SWT)~\cite{Bravyi2011,Winkler2003} with $\mathbf{k}_\parallel$ as the perturbation parameter. The contributions from all remote sub-bands, the $\mathcal{B}$ set, are eliminated up to the desired perturbation order. More explicitly, the unperturbed Hamiltonian is $\mathbf{E}_0$ ($\mathbf{k_\parallel}$-independent), while the perturbation $\mathbf{W}(\mathbf{k}_\parallel) = \mathbf{W}_1(\mathbf{k}_\parallel) + \mathbf{W}_2(\mathbf{k}_\parallel)$ is at least linear in $\mathbf{k}_\parallel$:

\begin{subequations}\label{Wpert}
\begin{align}
    \mathbf{W}_1(\mathbf{k}_\parallel) &=  i\alpha_0\mathbf{M}_1 k_- + \text{h.c.}, \\
    \mathbf{W}_2(\mathbf{k}_\parallel) &= \alpha_0\left[\mathbf{M}_\gamma k_\parallel^2 + \left(\mathbf{M}_2k_-^2  + \text{h.c.}\right)\right].
\end{align}    
\end{subequations}

\noindent We also define the following projection operators:

\begin{align}
    P &= \sum_{\alpha\in\mathcal{A}}{\ket{\varphi_\alpha}\bra{\varphi_\alpha}}, \\
    Q &= \sum_{\lambda\in\mathcal{B}}{\ket{\varphi_\lambda}\bra{\varphi_\lambda}},
\end{align}

\noindent and superoperators:

\begin{align}
    \mathcal{P}[X] &= PXP, \\
    \mathcal{Q}[X] &= QXQ, \\
    \mathcal{D}[X] &= PXP + QXQ, \\
    \mathcal{O}[X] &= PXQ + QXP, \\
    \mathcal{L}[X] &= \sum_{i,j}{\ket{\varphi_i}\frac{\bra{\varphi_i}\mathcal{O}[X]\ket{\varphi_j}}{E_i-E_j}\bra{\varphi_j}},
\end{align}

\noindent where $E_i$ is the energy of sub-band $i$ of the unperturbed Hamiltonian (or at $\mathbf{k}_\parallel = 0$). It is clear from the definition of $\mathcal{L}[X]$ that $\mathcal{O}[\mathcal{L}[X]] = \mathcal{L}[X]$ and $\mathcal{D}[\mathcal{L}[X]] = 0$. For the \SiGe{} material system, we apply perturbation theory up to $2$\textsuperscript{nd} order (and only keeping terms up to $k^2$). The effective Hamiltonian for the $\mathcal{A}$ set is given by

\begin{equation}
    {H}_\mathrm{eff}(\mathbf{k}_\parallel) = \mathcal{P}[\mathbf{E}_0] + \mathcal{P}[\mathbf{W}_1] + \mathcal{P}[\mathbf{W}_2] + \mathcal{P}\left[\mathrm{sw}_2(\mathbf{W}_1,\mathbf{W}_1)\right],
\end{equation}

\noindent where

\begin{equation}
    \mathrm{sw}_2(X,Y) = \frac{1}{2}\left(\mathcal{L}[X]Y - X\mathcal{L}[Y]\right).
\end{equation}

The second order term involves products of $\mathbf{W}_1$ with itself, and will ultimately result in terms that can be collected back into the $\mathbf{M}$ matrices that define the original Hamiltonian. Explicitly:

\begin{align*}
    \frac{\mathbf{W}_1\mathbf{W}_1}{\alpha_0^2} &= \left(i\mathbf{M}_1k_- - i\mathbf{M}_1^\dagger k_+\right)\left(i\mathbf{M}_1k_- - i\mathbf{M}_1^\dagger k_+\right) \\
    &= \left(\mathbf{M}_1\mathbf{M}_1^\dagger + \mathbf{M}_1^\dagger\mathbf{M}_1\right)k_\parallel^2 - \left(\mathbf{M}_1\mathbf{M}_1k_-^2 + \mathrm{h.c.}\right).
\end{align*}

\noindent By identification with the original Hamiltonian \eqref{Hk} we define the $\mathcal{A}$-set $\mathcal{M}$-matrices

\begin{subequations}
    \begin{align}
        \mathcal{M}_\gamma &= \mathcal{P}[\mathbf{M}_\gamma] + \alpha_0\mathcal{P}\left[\mathrm{sw}_2(\mathbf{M}_1,\mathbf{M}_1^\dagger) + \mathrm{sw}_2(\mathbf{M}_1^\dagger,\mathbf{M}_1)\right], \\
        \mathcal{M}_2 &= \mathcal{P}[\mathbf{M}_2] - \alpha_0\mathcal{P}\left[\mathrm{sw}_2(\mathbf{M}_1,\mathbf{M}_1)\right], \\
    \end{align}
\end{subequations}

\noindent and for completeness:

\begin{subequations}
    \begin{align}
        \mathcal{E}_0 &= \mathcal{P}[\mathbf{E}_0], \\
        \mathcal{M}_1 &= \mathcal{P}[\mathbf{M}_1].
    \end{align}
\end{subequations}

In the \GeSn{} material system, we apply perturbation theory up to $1$\textsuperscript{st} order. In this case the states belonging to the $\mathcal{B}$ set are simply discarded from the Hamiltonian, which becomes

\begin{equation}
    {H}_\mathrm{eff}(\mathbf{k}_\parallel) = \mathcal{P}[\mathbf{E}_0] + \mathcal{P}[\mathbf{W}_1] + \mathcal{P}[\mathbf{W}_2],
\end{equation}

\noindent and with corresponding $\mathcal{M}$-matrices

\begin{subequations}
    \begin{align}
        \mathcal{E}_0 &= \mathcal{P}[\mathbf{E}_0], \\
        \mathcal{M}_\gamma &= \mathcal{P}[\mathbf{M}_\gamma], \\
        \mathcal{M}_1 &= \mathcal{P}[\mathbf{M}_1], \\
        \mathcal{M}_2 &= \mathcal{P}[\mathbf{M}_2].
    \end{align}
\end{subequations}

In both material systems these matrices of reduced dimension can replace those in the original Hamiltonian \eqref{Hk} to get

\begin{equation}\label{Hk_}
    {H}_\mathrm{eff}(\mathbf{k}_\parallel) = \mathcal{E}_0 + \alpha_0\left[\mathcal{M}_\gamma k_\parallel^2 + \left(i\mathcal{M}_1 k_- + \mathcal{M}_2k_-^2 + \text{h.c.}\right)\right].
\end{equation}

\begin{figure}
    \centering
    \includegraphics[width=0.5\linewidth]{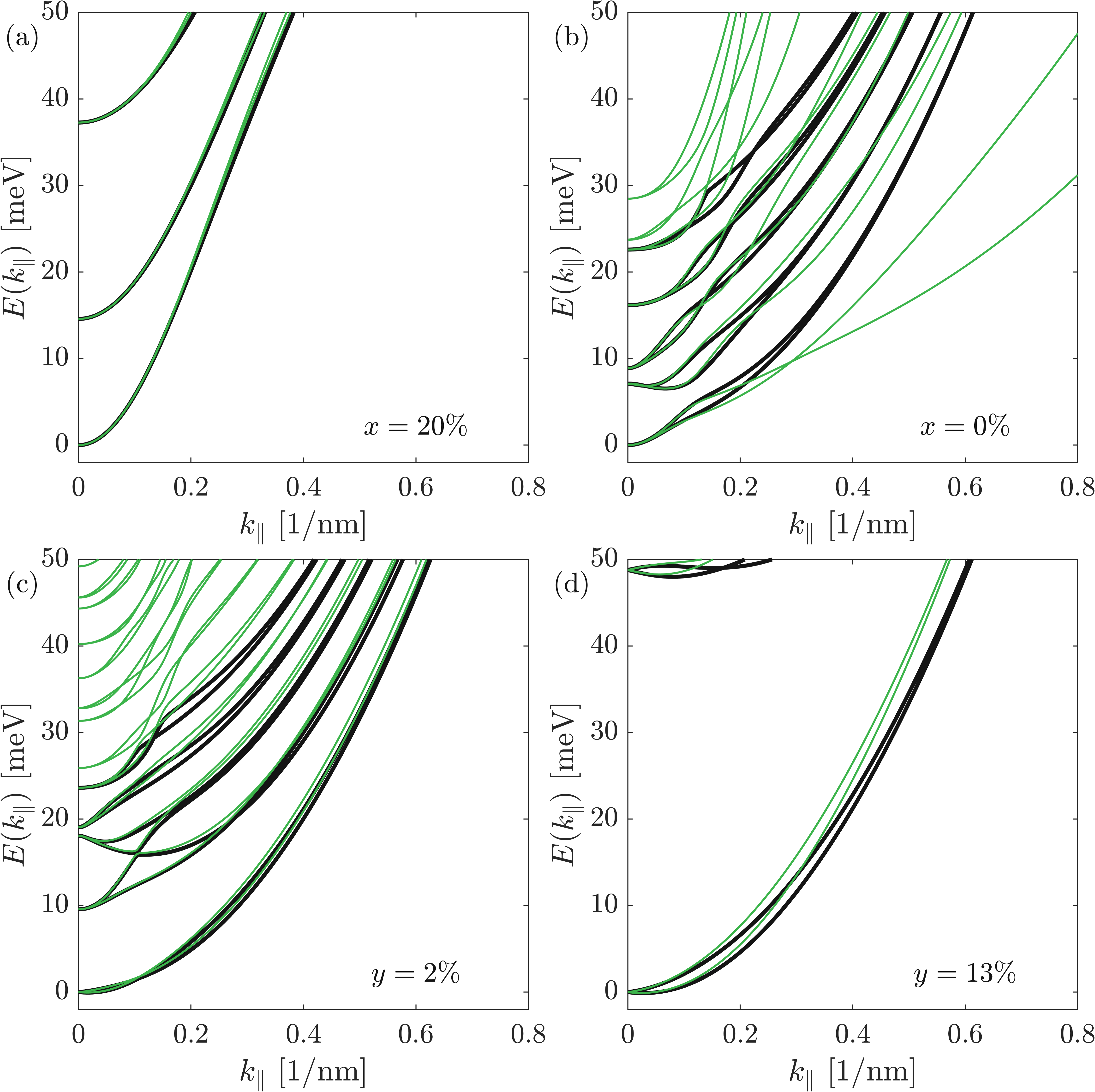}
    \caption{$E(k_\parallel)$ dispersion from the full Hamiltonian (black) computed from a $400$-dimensional Hilbert space and from the effective Hamiltonian $H_\mathrm{eff}$ (green). Top row: $2$\textsuperscript{nd} order SWT with $\mathcal{A} = \{\h{1},\h{2},\h{3},\h{4},\h{5},\e{1},\e{2}\}$. Bottom row: $1$\textsuperscript{st} order SWT with $\mathcal{A} = \{\h{1},\cdots,\h{20},\e{1},\cdots,\e{10}\}$. (a) Compressive strained Ge with $x = 20\%$. (b) Unstrained Ge. (c) Tensile strained Ge with $y = 2\%$. (d) Tensile strained Ge with $y = 13\%$. $E_z = 1\,\mathrm{mV}/\mathrm{nm}$.}
    \label{fig:si_Heff}
\end{figure}

Fig. \ref{fig:si_Heff} shows the energy dispersion of compressive strained Ge ($x = 20\%$), relaxed Ge ($x = 0\%$), slightly tensile strained Ge ($y = 2\%$) and highly tensile strained Ge ($y = 13\%$) computed from the diagonalization of the full $k\cdot p$ matrix $\mathbf{H}_\mathrm{Ge}$ compared to the dispersion provided by the effective Hamiltonian $H_\mathrm{eff}$. The dimension of the full Hilbert space is $2\cdot 200 = 400$.

\section{\texorpdfstring{Rashba spin splitting and Fermi velocities in Ge/\SiGe{} and Ge/\GeSn{}}{}}

\begin{figure}[th]
    \centering

    \includegraphics[width=\linewidth]{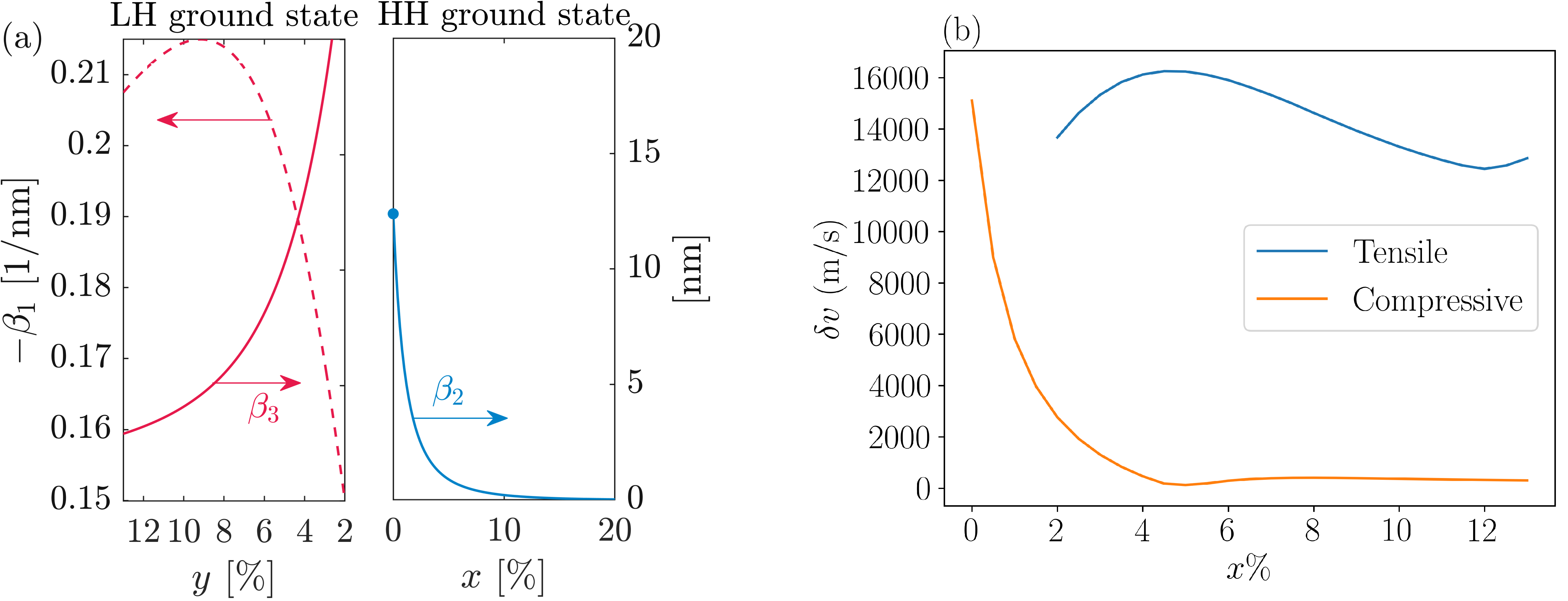}
    \caption{(a) Left panel: $-\beta_1$ (left $y$-axis, dashed line) and $\beta_3$ (right $y$-axis, solid line) as a function of the Sn content $y$. Right panel: $\beta_2$ as a function of the Si content $x$. The right $y$-axis in both panels share the same units and scale. $E_z = 1\,\mathrm{mV}/\mathrm{nm}$. (b) Maximum velocity difference against strain, for the two lowest bands of a germanium lead with $W=100$ nm. The maximum is taken in the $\mu$ range where only the first sub-band is occupied.}
    \label{fig:beta}

\end{figure}

In the limit of small wavenumbers $\mathbf{k}_\parallel$, the HH and the LH ground state Hamiltonian is well approximated by the following $2\times 2$ Hamiltonian:

\begin{equation}
    {H}_\mathrm{eff}^\pm = \frac{\hbar^2}{2m_0}\gamma k_\parallel^2 + \frac{\hbar^2}{2m_0}\left[i\left(\beta_1k_- - \beta_2k_+^3 + \beta_3k_\parallel^2k_-\right)\sigma_\pm + \text{h.c.}\right],
\end{equation}

\noindent where $\sigma_\pm = (\sigma_x\pm i\sigma_y)/2$ and $\pm$ is for LH or HH ground states, respectively, and $m_0/\gamma$ is the in-plane effective mass. This Hamiltonian is derived using the perturbative framework presented in Section \ref{sec:pert}, but using $3$\textsuperscript{rd} order perturbation theory~\cite{Winkler2003} and with $\mathcal{A}=\{\h{1}\}$ for the Ge/\SiGe{} heterostructure and with $\mathcal{A}=\{\e{1}\}$ for the Ge/\GeSn{} heterostructure. In the axial approximation ($\gamma_2\approx\gamma_3$) and neglecting the cubic-in-$J$ correction to the $g$-factor ($q\approx0$~\cite{Lawaetz1971}), $\beta_1 = 0$ and $\beta_3 = 0$ for HHs, while $\beta_2 = 0$ for LHs. The magnitude of the Rashba coefficients $\beta_i$ as a function of the Si content $x$ and the Sn content $y$ is shown in Fig. \ref{fig:beta}.

Perturbation theory provides explicit expressions for the Rashba parameters. Using \eqref{Wpert}, the linear-in-$k$ Rashba term is (for LHs only)

\begin{equation}
    \beta_1 = T^\upeta_{1,1} = 3\sqrt{2}\Im\{\bra{f_1^\circ}\{\tilde{\gamma},k_z\} + [\kappa,k_z]\ket{f_1^\perp}\}.
\end{equation}

The cubic-in-$k$ term $\beta_3$ is the $\sigma_+$ matrix element of the $2\times 2$ matrix (LHs only)

\begin{align}
    \boldsymbol{\upbeta}_3 &= \alpha_0\mathcal{P}\left[\mathrm{sw}_2\left(\mathbf{M}_\gamma,\mathbf{M}_1\right) -\mathrm{sw}_2\left(\mathbf{M}_1^\dagger,\mathbf{M}_2\right) + \text{c.p.}\right] +\alpha_0^2\mathcal{P}\left[\mathrm{sw}_3\left(\mathbf{M}_1,\mathbf{M}_1,\mathbf{M}_1^\dagger\right) + \text{c.p.}\right] \\
    &= \beta_3\sigma_+,
\end{align}

\noindent where c.p. means cyclic permutations of the arguments to the functions $\mathrm{sw}_n(\cdot)$ and

\begin{equation}
    \mathrm{sw}_3(X,Y,Z) = \frac{1}{2}\left(\mathcal{L}[\mathcal{L}[X]Y]Z + X\mathcal{L}[Y\mathcal{L}[Z]] + \mathcal{L}[XP\mathcal{L}[Y]]Z + X\mathcal{L}[\mathcal{L}[Y]PZ]\right).
\end{equation}

Finally, the cubic-in-$k$ term $\beta_2$ is the $\sigma_+$ matrix element of the $2\times 2$ matrix (HHs only)

\begin{align}
    \boldsymbol{\upbeta}_2 &= \alpha_0\mathcal{P}\left[\mathrm{sw}_2\left(\mathbf{M}_1,\mathbf{M}_2\right) + \text{c.p.}\right] - \alpha_0^2\mathcal{P}\left[\mathrm{sw}_3\left(\mathbf{M}_1,\mathbf{M}_1,\mathbf{M}_1\right)\right] \\
    &= \beta_2\sigma_+.
\end{align}

\section{Material parametrization}\label{sec:mtr}

The $k\cdot p$ framework described in section \ref{sec:kp} uses $11$ material parameters. Those are the lattice constant $a_0$, the valence band energy $\mathcal{E}_{\Gamma_5^+}$ (relative to vacuum; corresponds to $E_{v,\text{avg}}$ in Reference \cite{VandeWalle1989}), the spin-orbit gap $\Delta_0$ between the degenerate HH/LH bands and the split-off band, the three Luttinger parameters $\gamma_1$, $\gamma_2$, and $\gamma_3$, the hole $g$-factor parameter $\kappa$, the elastic constants $c_{11}$ and $c_{12}$, and the deformation potential constants $a_v$ and $b$. A list of all parameters appearing in the $k\cdot p$ framework for Si, Ge and Sn is presented in Table \ref{tb:parameters}.

All material parameters of the \SiGe{} and \GeSn{} alloys, except the effective masses and $g$-factors, were calculated in the full composition range by interpolating the parameters from the pure elements:

\begin{equation}\label{Vegard}
  P(x) = (1-x)P^\text{A} + xP^\text{B} - x(1-x)b_P.
\end{equation}

\begin{table}[th!]
  \centering
  \begin{threeparttable}
    \caption{Material parameters and bowings appearing in the $k\cdot p$ Hamiltonian.\tnote{ }}\label{tb:parameters}
    \begin{tabular}{l l c c c c c}
      \toprule
      & & {\bf Si} & \SiGe{} & {\bf Ge} & \GeSn{} & {\bf Sn} \\
      \midrule
      \multicolumn{7}{l}{Lattice constant} \\
      $a_0$ & [\AA\ @ $0\,\text{K}$] & $5.429815$\tnote{a} & $0.0246$\tnote{g} & $5.652357$\tnote{a} & $-0.083$\tnote{h} & $6.480117$\tnote{b} \\
      \hline
      \multicolumn{7}{l}{Band energies} \\
      $\Delta_0$ & [eV] & $0.0441$\tnote{b} & & $0.290$\tnote{b} & $-0.100$\tnote{h} & $0.770$\tnote{f} \\
      $\mathcal{E}_{\Gamma_5^+}$ & [eV] & $-0.68$\tnote{c$\dag$} & & $0$ & & $0.69$\tnote{i$\dag$} \\
      \hline
      \multicolumn{7}{l}{Mechanical properties} \\
      $c_{11}$ & [GPa] & $165.77$\tnote{b} & & $124$\tnote{b} & & $69.0$\tnote{b} \\
      $c_{12}$ & [GPa] & $63.93$\tnote{b} & & $41.3$\tnote{b} & & $29.3$\tnote{b} \\
      \hline
      \multicolumn{7}{l}{Deformation potentials} \\
      $a_v$ & [eV] & $2.46$\tnote{c} & & $1.24$\tnote{c} & & $1.58$\tnote{j} \\
      $b$ & [eV] & $-2.1$\tnote{d} & & $-2.86$\tnote{d} & & $-2.7$\tnote{k} \\
      \hline
      \multicolumn{7}{l}{Effective mass and spin parameters} \\
      $\gamma_1$ & . & $4.285$\tnote{e} & \tnote{\P} & $13.38$\tnote{e} & \tnote{\#} &  \\
      $\gamma_2$ & . & $0.339$\tnote{e} & \tnote{\P} & $4.24$\tnote{e} & \tnote{\#} &  \\
      $\gamma_3$ & . & $1.446$\tnote{e} & \tnote{\P} & $5.69$\tnote{e} & \tnote{\#} &  \\
      $\kappa$ & . & $-0.26$\tnote{f} & \tnote{\P} & $3.41$\tnote{f} & & $-11.84$\tnote{f} \\
      \bottomrule
    \end{tabular}
    \begin{tablenotes}
    \item [ ] References: $^\text{a}$:\cite{Reeber1996}, $^\text{b}$:\cite{Madelung1991}, $^\text{c}$:\cite{VandeWalle1989}, $^\text{d}$:\cite{VandeWalle1986}, $^\text{e}$:\cite{Winkler2003}, $^\text{f}$:\cite{Lawaetz1971}, $^\text{g}$:\cite{Dismukes1964}, $^\text{h}$:\cite{Polak2017}, $^\text{i}$:\cite{Menendez2004}, $^\text{j}$:\cite{Brudevoll1993}, $^\text{k}$:\cite{Willatzen1995}
    \item [$\dag$] Relative to Ge
    \item [\P] See References \onlinecite{Winkler1996,Lawaetz1971}
    \item [\#] See Equation \eqref{gamma}
    \end{tablenotes}
  \end{threeparttable}
\end{table}

Here $0 \leq x \leq 1$ is the alloy fraction and $b_P$ is a bowing constant for the parameter $P$, if necessary. In the \SiGe{} material system, the three Luttinger parameters $\gamma_{1,2,3}$ and the hole $g$-factor parameter $\kappa$ were computed based on the approach employed in Reference \cite{Winkler1996} and following the theory outlined in Reference \cite{Lawaetz1971}.

In the \GeSn{} material system, the three Luttinger parameters $\gamma_{1,2,3}$ were interpolated between pure Ge and Ge$_{0.80}$Sn$_{0.20}$ using the data from Reference \cite{LuLow2012}, giving

\begin{equation}\label{gamma}
    \gamma_i(y) = \gamma_i^\mathrm{Ge}\left(1 - \frac{y}{0.2}\right) + \gamma_i^\mathrm{GeSn}\left(\frac{y}{0.2}\right) - b_i\frac{y}{0.2}\left(1-\frac{y}{0.2}\right),
\end{equation}

with $\gamma_i^\mathrm{Ge}$ listed in table \ref{tb:parameters}, and $\gamma_1^\mathrm{GeSn} = 29.2108$, $\gamma_2^\mathrm{GeSn} = 12.2413$, $\gamma_3^\mathrm{GeSn} = 13.7387$ and $b_1 = 20.3391$, $b_2 = 9.6609$, $b_3 = 9.8187$.

\section{Modeling the Germanium Josephson junction}
\subsection{Numerical model}

\begin{figure}[h!]
    \centering
    \includegraphics[width=0.5\linewidth]{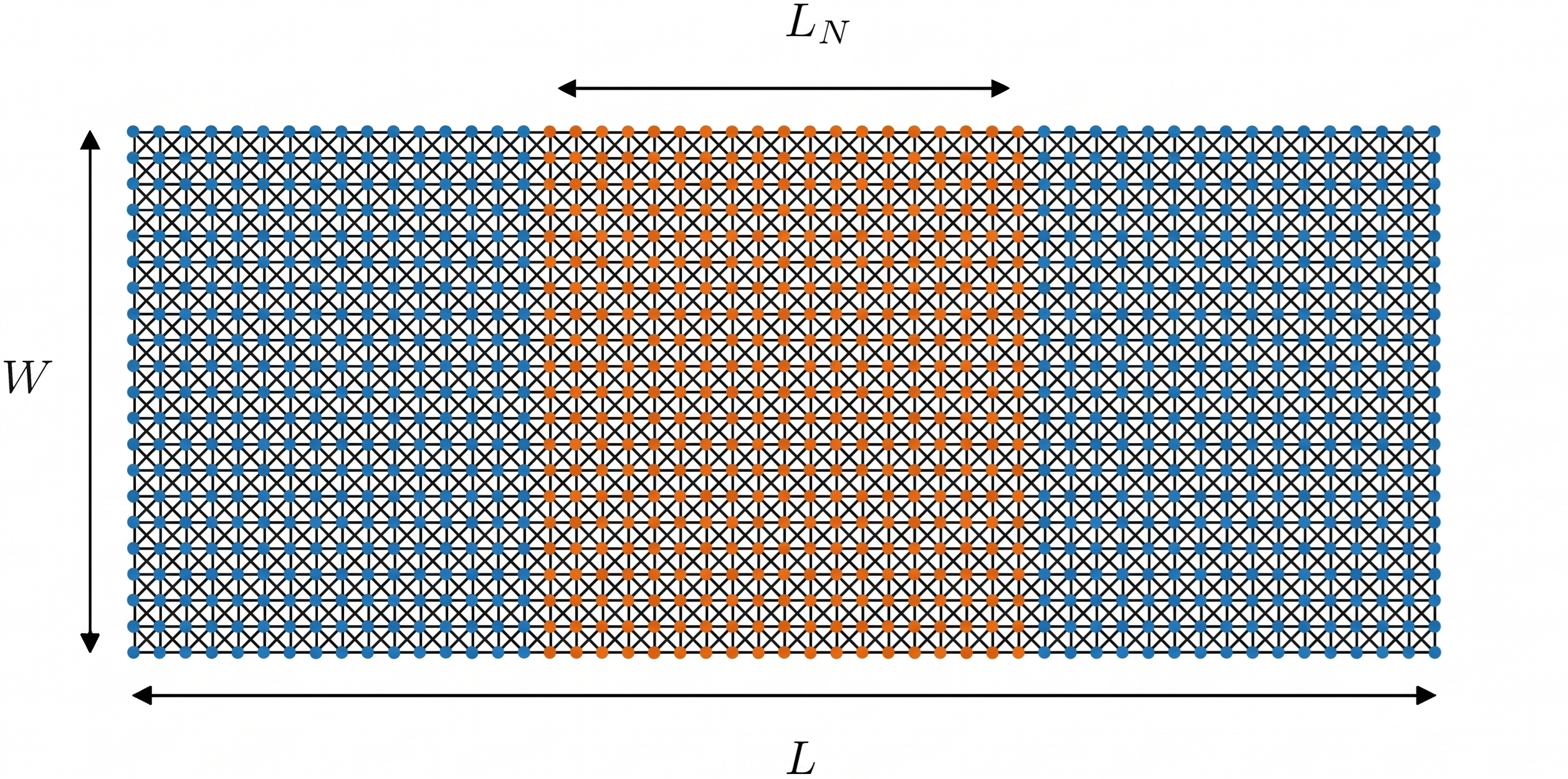}
    \caption{Sketch of a germanium junction. Each dot is a site and each line represents an hopping term. The orange area is the normal region while the blue areas are the superconducting regions.}
    \label{fig:jj_sketch}
\end{figure}

To find the ABS spectrum, we consider a finite 2D junction of total length $L$ (along $x$), width $W$ (along $y$), and length of the normal region $L_N$, as shown in Fig.~\ref{fig:jj_sketch}. To model the junction, we discretize the effective Hamiltonian in Eq. (3) of the main text on a lattice with lattice constant $a=3$ nm
, with a space dependent s-wave coupling $\Delta(x)$
\begin{equation}
    \Delta(x) = \left\{
    \begin{array}{ll}
        \Delta e^{i\phi/2} & -L/2\leq x<-L_N/2\quad \mbox{(left side of the junction)}\\
        & \\
        0 & |x|\leq L_N/2 \quad \mbox{(normal region)}\\
        & \\
        \Delta e^{-i\phi/2} & L_N/2<x\leq L/2 \quad \mbox{(right side of the junction)}
    \end{array}
    \right.
\end{equation}
To choose the value of the total length $L$, we use the following scheme:
\begin{enumerate}
    \item We compute the range of value of the chemical potential $\mu$ so that we have only one well-separated Andreev doublet in the junction. To this end, we compute the band structure of a normal Ge \footnote{To model the normal Ge, we can discretize the hamiltonian in Eq. (2) or the hamiltonian in Eq. (3) of the main text with $\Delta=0$.} lead (assuming translational invariance along $x$) of width $W$. We take $\mu$ within the bottom of the 1st sub-band and the bottom of the 2nd sub-band.
    \item Within the previous range of chemical potential, we compute the value of $\mu$ for which the coherence length $\xi$ of the evanescent modes in a superconductive lead of width $W$, is equal to some fixed value (we choose $\xi = 100,\, 125,\, 150, \, 175,\, 200$ nm)
    \item We choose the length of both superconductive regions of the junction to be equal to $2\xi$, obtaining a total length of $L=2 \cdot 2\xi + L_N$. This guarantees that the system numerically diagonalized is large enough to accommodate an Andreev bound state wave function with decay length $\approx \xi$ in the superconducting leads, without substantial finite-size effects.
\end{enumerate}
From the band structure of the normal Ge lead, we also extract Fermi velocities $v_F$ for the values of $\mu$ computed using the previous scheme.

Depending on the chosen values of $L_N, \, W$ and $\xi$, the diagonalization of the hamiltonian of the whole junction can be quite memory demanding. Thus, due to the limited memory resources available for the calculations, the $W$ range in Fig. 2(c) of the main text decreases for larger values of $\xi$.

\subsection{Semi-analytical model}

\begin{figure}[h!]
    \centering
    \includegraphics[width=0.35\linewidth]{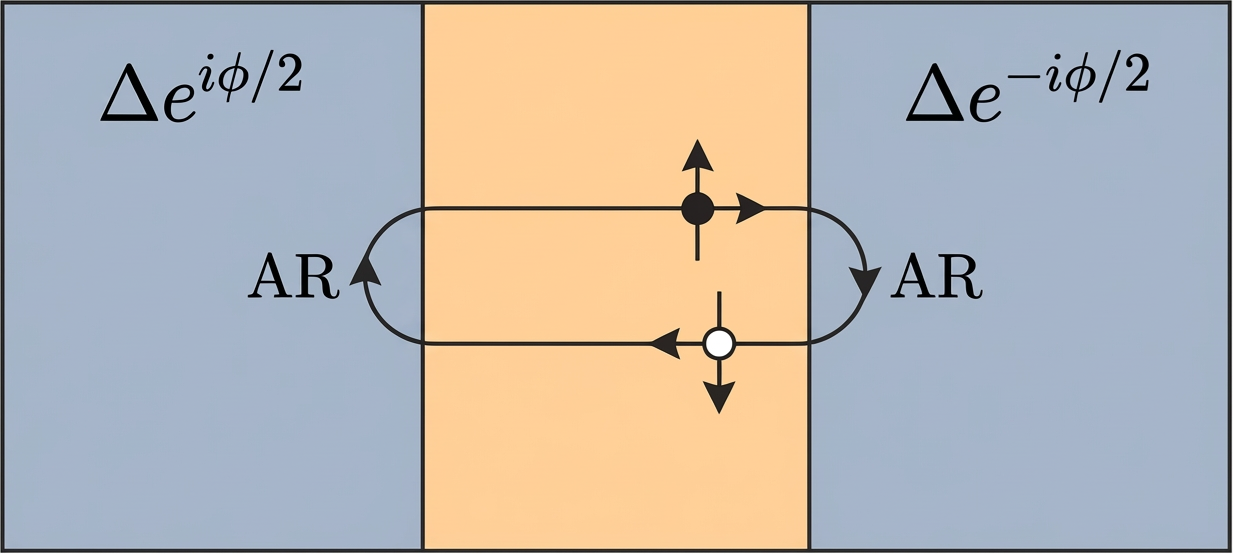}
    \caption{Sketch of a closed trajectory of an Andreev bound state.}
    \label{fig:ABS}
\end{figure}

To find the energies of Andreev bound states in a long junction, we can apply the Bohr-Sommerfeld quantization rule by imposing that the total phase acquired by a quasi-particle along a close trajectory is $2n\pi$, with $n$ integer. This condition ensures that the wave function is single-valued in every point in the space. As shown in Figure \ref{fig:ABS}, we can compute the total phase as a sum of 4 terms:
\begin{align}
    \begin{array}{ccccccccc}
      \longrightarrow   & +&  \supset & +& \longleftarrow & +& \subset & =& 2n\pi\\
      &&&&&&&& \\
      k_e L_N & +& -\arccos\left(\frac{E}{\Delta}\right)-\left(-\frac{\varphi}{2}\right) & +& -k_h L_N & +& -\arccos\left(\frac{E}{\Delta}\right)+\left(\frac{\varphi}{2}\right) & =& 2n\pi\,.
    \end{array}
\end{align}   
In the end, we obtain the following
\begin{equation} \label{eq:bohr-somm}
    \varphi -2\arccos\left(\frac{E}{\Delta}\right) + (k_e-k_h) L_N=2n\pi\,.
\end{equation}
By assuming a linear dispersion around $k_F$, one can make the energy dependence of the third term of the left-hand side explicit
\begin{equation}
    E = \frac{\hbar^2 k^2}{2m} - \mu = \frac{\hbar^2 k^2}{2m} - \frac{\hbar^2 k_F^2}{2m} \rightarrow k = \sqrt{k_F^2+ \frac{2mE}{\hbar^2}} \approx k_F^2 + \frac{mE}{\hbar^2k_F^2}= k_F^2 + \frac{E}{\hbar v_F}
\end{equation}
For holes, we obtain
\begin{equation}
    E \approx k_F^2 - \frac{E}{\hbar v_F}
\end{equation}
By substitution in equation \eqref{eq:bohr-somm}
\begin{equation} \label{eq:bohr-somm_final}
    \varphi -2\arccos\left(\frac{E}{\Delta}\right) + \frac{2E}{\hbar v_F} L_N=2n\pi
\end{equation}
The presence of $v_F$ in the last equation makes clear that spin-splitting of the energy levels is expected if the spin-orbit interaction is present.
We take $n=0$ and solve equation \eqref{eq:bohr-somm_final} numerically to find the energies of the first Andreev doublet.

\subsection{Matrix elements $\partial H/\partial \phi$}

We verify the possibility of a flux drive of the transition between the two spin states of the Andreev doublet, by computing the corresponding matrix $\partial H/\partial \phi$.
In accordance with Ref.~\cite{PhysRevResearch.3.013036} we find that, due to symmetry, the transition is forbidden for a symmetric system along $y$, thus we add an anti-symmetric transverse potential $E_y y$, with $E_y=0.03$ meV/nm, to activate the transition.

\begin{figure}[h!]
    \centering
    \includegraphics[width=0.35\linewidth]{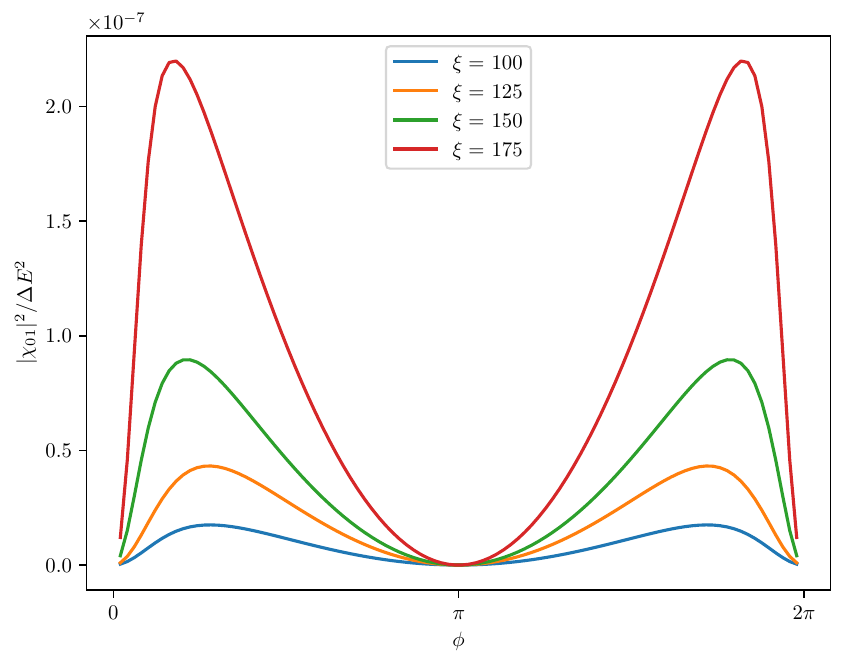}
    \caption{Phase dependence of the off-diagonal elements of $\partial H/\partial \phi$.}
    \label{fig:dHdphi}
\end{figure}
The matrix element has a maximum in correspondence of the maximum of the spin-splitting. 
However, compared to the spin-splitting, the matrix element is very small and the resulting Rabi frequencies would be much smaller than those obtained in the main text for electric driving.

\bibliography{references}